\def\mode{lncs}
\def\lncs{
	\documentclass[runningheads,orivec]{./llncs}		
	\setlength{\paperheight}{232.8mm}
	\setlength{\paperwidth}{151.5mm}
	\setlength\voffset     {-19mm}
	\setlength\hoffset     {-34mm}
	\bibliographystyle{./splncs04}
	\usepackage[firstpage]{draftwatermark}
	\SetWatermarkAngle{0}
	\SetWatermarkFontSize{10pt}
	\SetWatermarkHorCenter{153mm}
	\SetWatermarkVerCenter{240mm}
	\SetWatermarkLightness{0.6}
	\SetWatermarkText{Postprint{,} \monthyeardate\today}
}
\def\article{%
	\documentclass{article}	 
	\setlength{\paperheight} {232.8mm}
	\setlength{\paperwidth}  {151.5mm}
	\setlength\voffset       {-26mm}
	\setlength\hoffset       {-32mm}
	\bibliographystyle{plain}
	\usepackage[firstpage]{draftwatermark}
	\SetWatermarkAngle{0}
	\SetWatermarkFontSize{10pt}
	\SetWatermarkHorCenter{153mm}
	\SetWatermarkVerCenter{247mm}
	\SetWatermarkLightness{0.6}
	\SetWatermarkText{Draft{,} {\monthyeardate\today}}
}
\csname\mode\endcsname

\newif\ifshowtodos
\showtodostrue		

\newcommand{\authorandrei}	{Andrei~Tour}
\newcommand{\authorartem}	{Artem~Polyvyanyy}
\newcommand{\authoranna}	{Anna~Kalenkova}
\newcommand{\authorarik}	{Arik~Senderovich}

\newcommand{\articletitle}   {Agent Miner: An Algorithm for Discovering Agent Systems from Event Data}
\newcommand{\articlesubjet}  {Computer Science, Process Mining, Agent System Mining}
\newcommand{\articleauthors} {\authorandrei, \authorartem, \authoranna, and \authorarik}
\usepackage{booktabs}
\usepackage{multirow}
\usepackage{nicefrac}
\usepackage[table,xcdraw]{xcolor}
\usepackage[algoruled,vlined,linesnumbered]{algorithm2e}
\usepackage{amsmath}
\usepackage{amsfonts}
\usepackage{mdframed}
\usepackage{paralist}
\setdefaultleftmargin{1em}{1em}{1em}{1em}{1em}{1em}
\usepackage{datetime}
\newdateformat{monthyeardate}{\monthname[\THEMONTH] \THEYEAR}
\usepackage[pdfsubject={\articlesubjet},
pdftitle={\articletitle},
pdfauthor={\articleauthors},
pdfkeywords={\articlesubjet},
colorlinks=false,
pdfborder={0 0 0}]{hyperref}

\hypersetup{bookmarksdepth}
\usepackage[algoruled]{algorithm2e}
\ifshowtodos
\usepackage[textsize=scriptsize,colorinlistoftodos]{todonotes}
\else
\usepackage[textsize=scriptsize,colorinlistoftodos,disable]{todonotes} 
\fi
\let\todonote\todo
\renewcommand{\todo}[2]{\todonote[inline,color=red!20]{TODO (#1): #2}}

\usepackage{tikz,pgf,xcolor}
\usetikzlibrary{arrows,automata,trees,plotmarks,shadows,shapes,decorations.pathmorphing,backgrounds,positioning,fit,petri}
\usepackage{pgfplots}
\usetikzlibrary{pgfplots.groupplots}
\usepackage{url}
\usepackage{mathtools} 
\usepackage[capitalise,nameinlink]{cleveref}

\definecolor{mybluecolor}{RGB}{50,106,218}
\definecolor{myredcolor}{RGB}{176,53,53}
\definecolor{mygreencolor}{RGB}{93,172,0}
\definecolor{myyellowcolor}{RGB}{255,163,34}
\definecolor{mypurplecolor}{RGB}{86,35,132}
\definecolor{mytealcolor}{RGB}{30,161,165}

\newcommand{\splitatcommas}[1]{%
	\begingroup
	\ifnum\mathcode`,="8000
	\else
	\begingroup\lccode`~=`, \lowercase{\endgroup
		\edef~{\mathchar\the\mathcode`, \penalty0 \noexpand\hspace{0pt plus 3em}}%
	}\mathcode`,="8000
	\fi
	#1%
	\endgroup
}

\newcommand{\func}[3]{{{#1}:{#2} \rightarrow {#3}}}
\newcommand{\funcCall}[2]{{\ensuremath {\mathit{#1}}_{\!}\left({#2}\right)}}

\newcommand{\intintervalcc}[2]{{\ensuremath \left[#1 \,..\, #2\right]}}

\providecommand{\cardinality}[1]{\ensuremath \left|{#1}\right|}
\renewcommand{\cardinality}[1]{\ensuremath \left|{#1}\right|}

\newcommand{\mset}[1] {\ensuremath [\splitatcommas{#1}]}

\newcommand{\set}[1]{\ensuremath \left\{\splitatcommas{#1}\right\}}
\newcommand{\setbuilder}[2]{\ensuremath \left\{ #1 \;|\; #2 \right\}}

\newcommand{\sequence}[1]{\ensuremath \langle\splitatcommas{#1}\rangle}
\newcommand{\seqLength}[1]{\ensuremath \left|{#1}\right|}

\newcommand{\pair}[2]{\ensuremath \left(\splitatcommas{#1, #2}\right)}
\newcommand{\tuple}[1]{\ensuremath \left(\splitatcommas{#1}\right)}

\newcommand{\enabled}[3]{\ensuremath \pair{#1}{#2}\left[#3\right\rangle}
\newcommand{\occurrence}[4]{\ensuremath \pair{#1}{#2}\left[#3\right\rangle\pair{#1}{#4}}

\newcommand{\predicate}[3]{\ensuremath #1 \, #2 : #3}

\newcommand\fixithelp[2]{%
	\wd0=\dimexpr\wd0-\linewidth\relax%
	\ifdim\wd0>0pt\relax%
	\fixithelp{#1}{#2}%
	\else%
	\wd0=\dimexpr\wd0+\linewidth\relax
	\ifdim\wd0>.9\linewidth\relax%
	{\parfillskip0pt\relax#2\par}%
	\else%
	\ifdim\wd0>.8\linewidth\relax%
	{\parfillskip0pt\relax#2\hspace{.2\linewidth}\par}%
	\else%
	\ifdim\wd0<#1\linewidth\relax%
	{\parfillskip0pt\relax#2\par}%
	\else%
	\ifdim\wd0<.2\linewidth\relax%
	{\parfillskip0pt\relax#2\hspace{.8\linewidth}\mbox{}\par}%
	\else%
	#2%
	\fi
	\fi
	\fi
	\fi
	\fi%
}

\newcommand{\orcidartem}		{\href{https://orcid.org/0000-0002-7672-1643}{\protect\includegraphics[scale=0.05]{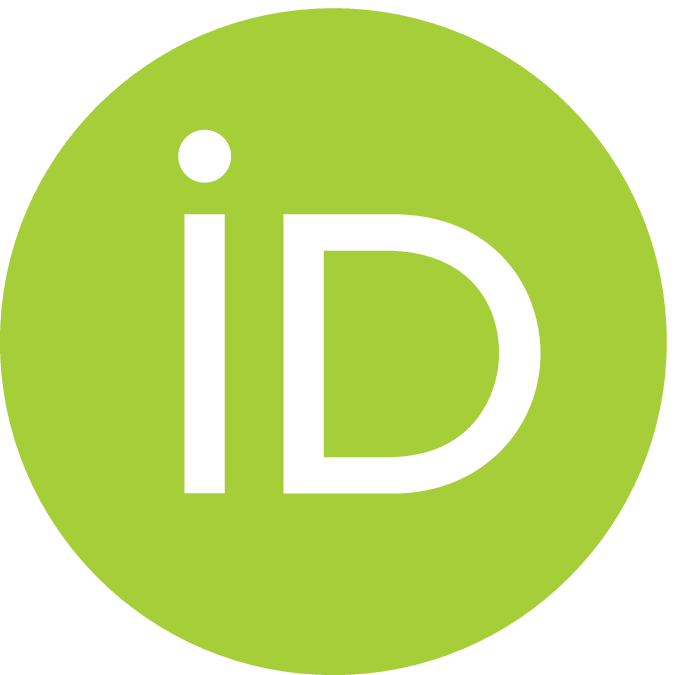}}}	
\newcommand{\orcidanna}			{\href{https://orcid.org/0000-0002-5088-7602}{\protect\includegraphics[scale=0.05]{orcid.eps}}} 
\newcommand{\orcidandrei}			{\href{https://orcid.org/0000-0002-1169-0866}{\protect\includegraphics[scale=0.05]{orcid.eps}}} 
\newcommand{\orcidarik}			{\href{https://orcid.org/0000-0003-4728-8024}{\protect\includegraphics[scale=0.05]{orcid.eps}}} 

\newcommand{\ie}					{i.e.,~}

\newtheorem{mytheorem}		{Theorem}
\newtheorem{mydefinition}	{Definition}
\newtheorem{mylemma}			{Lemma}
\newtheorem{myproposition}{Proposition}
\newtheorem{mycorollary}	{Corollary}
\newtheorem{myexample}		{Example}
\newtheorem{myconjecture}	{Conjecture}

\newtheorem{myinvariant}	{Invariant}

\numberwithin{mytheorem}		{section}
\numberwithin{mydefinition}	{section}
\numberwithin{mylemma}			{section}
\numberwithin{myproposition}{section}
\numberwithin{mycorollary}	{section}
\numberwithin{myexample}		{section}
\numberwithin{myconjecture}	{section}
\numberwithin{myremark}			{section}
\numberwithin{myinvariant}	{section}

\newenvironment{define}[3][]
{\begin{mydefinition}[#2]\label{#3}#1\normalfont}
	{\hfill\ensuremath{\lrcorner}\end{mydefinition}}

\addtocounter{mytheorem}{0}

\usepackage{caption}
\usepackage{subcaption}
\usepackage{algorithm2e}
\usepackage{comment}
\usepackage{wrapfig,booktabs}
\usepackage{booktabs}
\usepackage{multirow}

\let\oldmaketitle\maketitle
\renewcommand{\maketitle}{\oldmaketitle\setcounter{footnote}{0}}

\title{\articletitle}

\date{\today}

\begin{document} 
	
	\author{\authorandrei\inst{1}~\orcidandrei \and \authorartem\inst{1}~\orcidartem \and \authoranna\inst{2}~\orcidanna \and \authorarik\inst{3}~\orcidarik}
	
	\authorrunning{A. Tour, A. Polyvyanyy, A. Kalenkova, A. Senderovich}
	
	\institute{%
		The University of Melbourne, VIC, 3010, Australia\\
		\email{atour@student.unimelb.edu.au}; \email{artem.polyvyanyy@unimelb.edu.au}%
		\and%
		The University of Adelaide, Adelaide, Australia\\
		\email{anna.kalenkova@adelaide.edu.au}%
		\and%
		York University, Toronto, Ontario, Canada\\
		\email{sariks@yorku.ca}%
	}
	
	\maketitle
	
	\vspace{-3mm}
\begin{abstract}
	Process discovery studies ways to use event data generated by business processes and recorded by IT systems to construct models that describe the processes.
	Existing discovery algorithms are predominantly concerned with constructing process models that represent the control flow of the processes.
	Agent system mining argues that business processes often emerge from interactions of autonomous agents and uses event data to construct models of the agents and their interactions.
	This paper presents and evaluates Agent Miner, an algorithm for discovering models of agents and their interactions from event data composing the system that has executed the processes which generated the input data.
	The conducted evaluation using our open-source implementation of Agent Miner and publicly available industrial datasets confirms that our algorithm can provide insights into the process participants and their interaction patterns and often discovers models that describe the business processes more faithfully than process models discovered using conventional process discovery algorithms.
\end{abstract}

	\vspace{-5mm}
\vspace{-3mm}
\section{Introduction}
\label{sec:introduction}
\vspace{-2mm}


Process discovery is a subarea of process mining that studies ways to construct models that faithfully describe processes of a system based on event data the system has generated~\cite{Aalst2016}.
Constructed models assist analysts in understanding the system and, consecutively, deciding how to improve it.
The state-of-the-art process discovery algorithms build models that describe the control flow of the processes.
This focus on control flow has at least two limitations.
Firstly, the resulting models are not well suited for analyzing the behavior of individual process participants and their interactions, as activities and interactions performed by a specific actor are often scattered across a discovered control flow model.
Secondly, control flow models discovered from large data arrays are often too complex, the phenomenon known as \emph{spaghetti models}~\cite{Aalst2016}.
Models of interacting agents, or \emph{agent systems}, where agents are the process participants, address the former limitation by their very definition and, importantly, do \emph{not} necessarily grow in complexity with the growth of the amount of data they represent~\cite{Wolfram1984}.


Agent system mining is a type of process mining that studies ways to derive and use knowledge about systems composed of interacting agents based on the events these agents generate~\cite{tour2021agent}.
This paper presents and evaluates an agent system discovery algorithm.
Concretely, this paper makes these contributions:
\smallskip
\begin{compactitem}
	\item 
	It presents Agent Miner, a divide-and-conquer algorithm for discovering models of agents and their interactions from event data.
	The algorithm ``divides'' the input collection of events into several special subsets and ``conquers'' these subsets using conventional discovery algorithms, like Inductive Miner~\cite{leemans2013discovering} and Split Miner~\cite{Augusto2019}, to construct an agent system that has generated the data.
	\item
	It presents the results of an evaluation of Agent Miner based on our open-source implementation of the algorithm that shows that Agent Miner can discover agent and interaction models from publicly available industrial event data that i) provide an additional perspective on the system that has generated the data and ii) often describe the processes more faithfully than the corresponding process models constructed using conventional control flow discovery algorithms; thus, addressing the two stated limitations.
	\item
	It demonstrates the value of agent system mining and invites the community to more intensive explorations of the agent-based paradigm in process mining.
\end{compactitem}
\smallskip
The remainder of this paper proceeds as follows.
The next section presents a motivating example. 
Then, \Cref{sec:preliminaries} gives basic notions required for understanding the subsequent discussions.
\Cref{sec:agent_miner} presents our new discovery algorithm, while \cref{sec:eval} discusses the results of its evaluation.
Finally, \cref{sec:related:work} surveys related work before \cref{sec:conclusion} gives concluding remarks on this work.

\vspace{-3mm}
\section{Motivating Example}
\label{sec:motivating_example}
\vspace{-2mm}

\begin{figure}[b]
	\vspace{-6mm}
	\centering
	\includegraphics[width=0.75\textwidth]{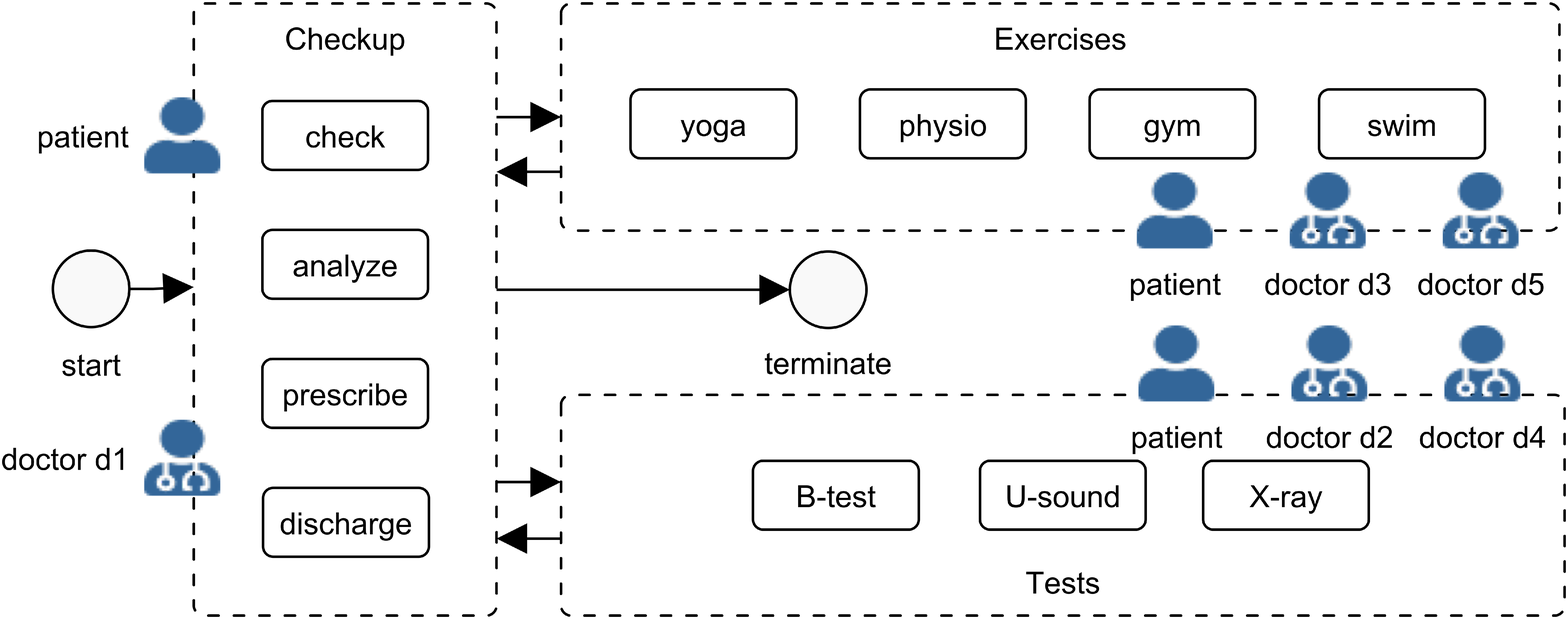}
	\vspace{-4mm}
	\caption{\small{A schematic visualization of the health surveillance process.}}
	\label{fig:running_example_viz}
	\vspace{-2mm}
\end{figure}

As a motivating example, we propose a hypothetical health surveillance process sketched in~\cref{fig:running_example_viz}.
The process starts with a check (see the \emph{check} activity in the figure) of the patient by doctor \textsl{d1}.
If the analysis (\emph{analyze}) of the results of the check indicates a risk of developing ill-health, the doctor prescribes (\emph{prescribe}) medical tests and preventive therapy.
The possible tests are blood test (\emph{B-test}), ultrasound (\emph{U-sound}), and \emph{X-ray}.
The therapy includes \emph{yoga}, \emph{physio}, \emph{gym}, and swimming (\emph{swim}).
The tests and the therapy are performed independently; any subset of tests and exercises can be prescribed.
Doctors \textsl{d2} and \textsl{d4} perform the tests, while the therapy is conducted by doctors \textsl{d3} and \textsl{d5}. 
Once the tests and the therapy are completed, doctor \textsl{d1} rechecks (\emph{check}) the patient.
If the new check shows good results, the patient is discharged (\emph{discharge}), and the process terminates; otherwise, further tests and therapy are prescribed. 
The information system that supports the process recorded a \emph{log} of events that stem from managing 1\,024 patients.
Each recorded event has four attributes that specify the \emph{timestamp} of the event occurrence, \emph{activity} that triggered the event, the patient \emph{case} the event relates to, and the doctor, or \emph{agent}, that performed the activity.

\begin{figure}[t]
	\vspace{-2mm}
	\centering
	\begin{subfigure}[b]{0.2\textwidth}
		\centering
		\includegraphics[height=3.3cm]{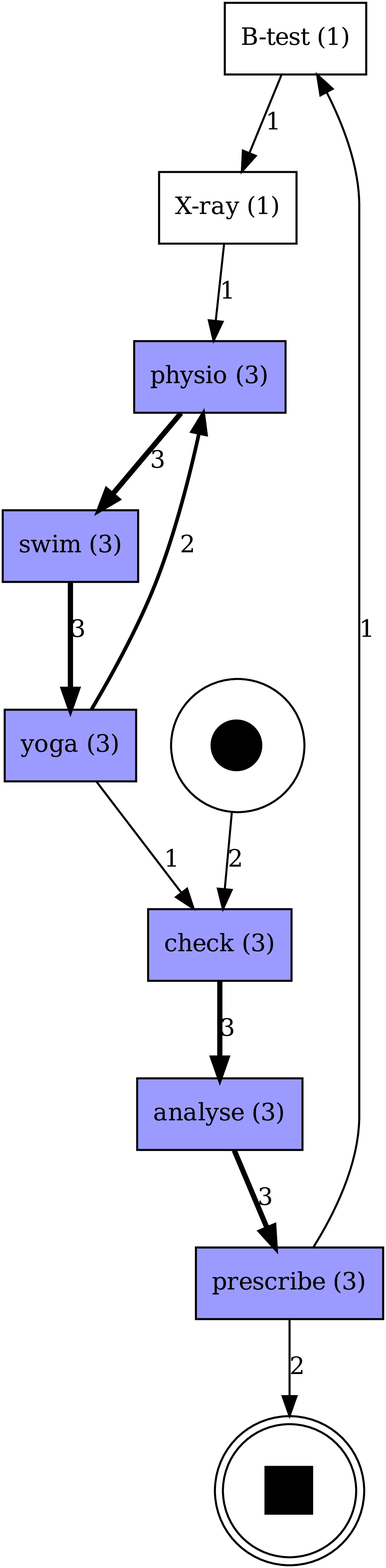}
		\caption{\small{2 traces}}
		\label{fig:dfg_2cases}
	\end{subfigure}
	\begin{subfigure}[b]{0.3\textwidth}
		\centering
		\includegraphics[height=3.3cm]{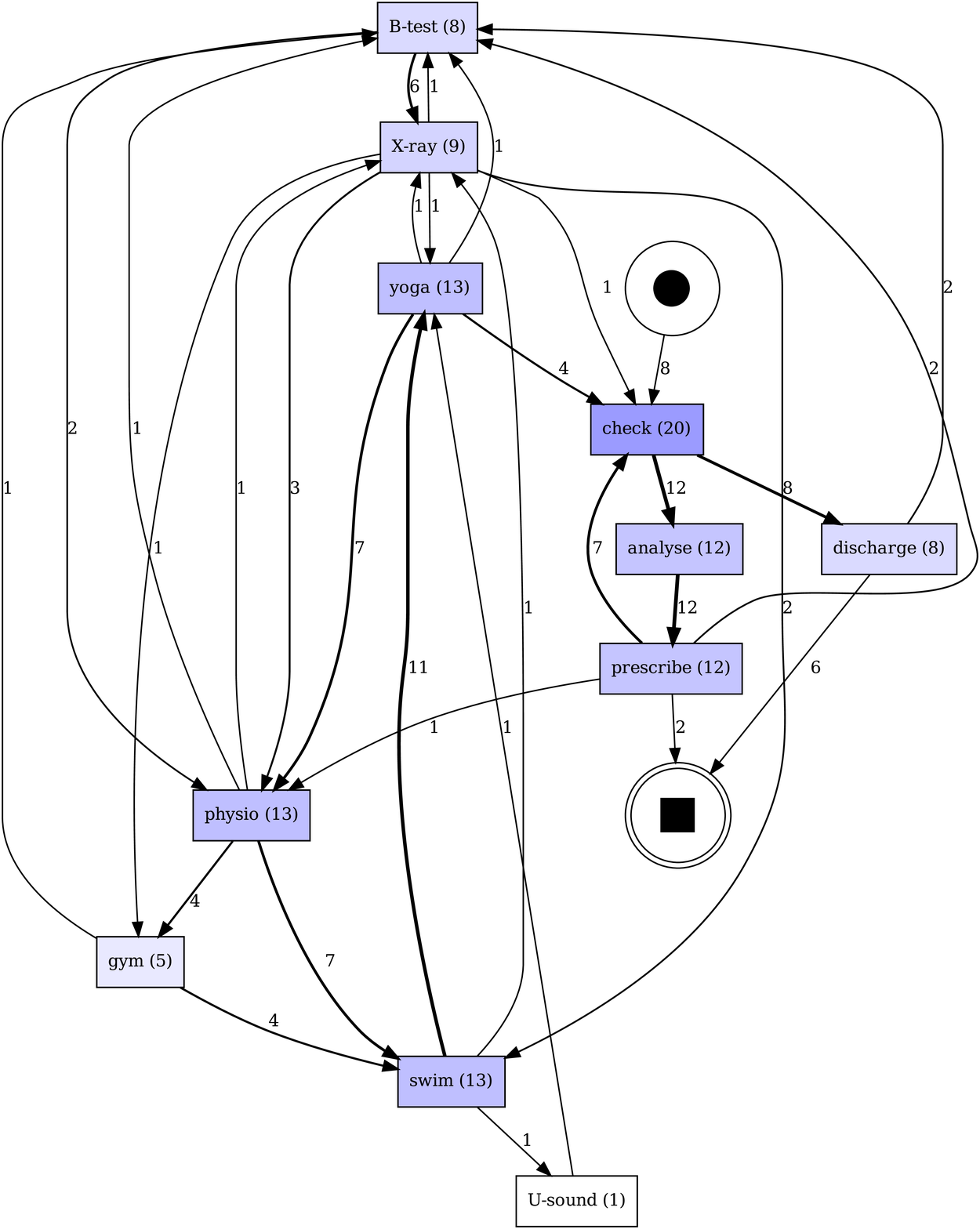}
		\caption{\small{8 traces}}
		\label{fig:dfg_8cases}
	\end{subfigure}
	\begin{subfigure}[b]{0.4\textwidth}
		\centering
		\includegraphics[height=3.3cm]{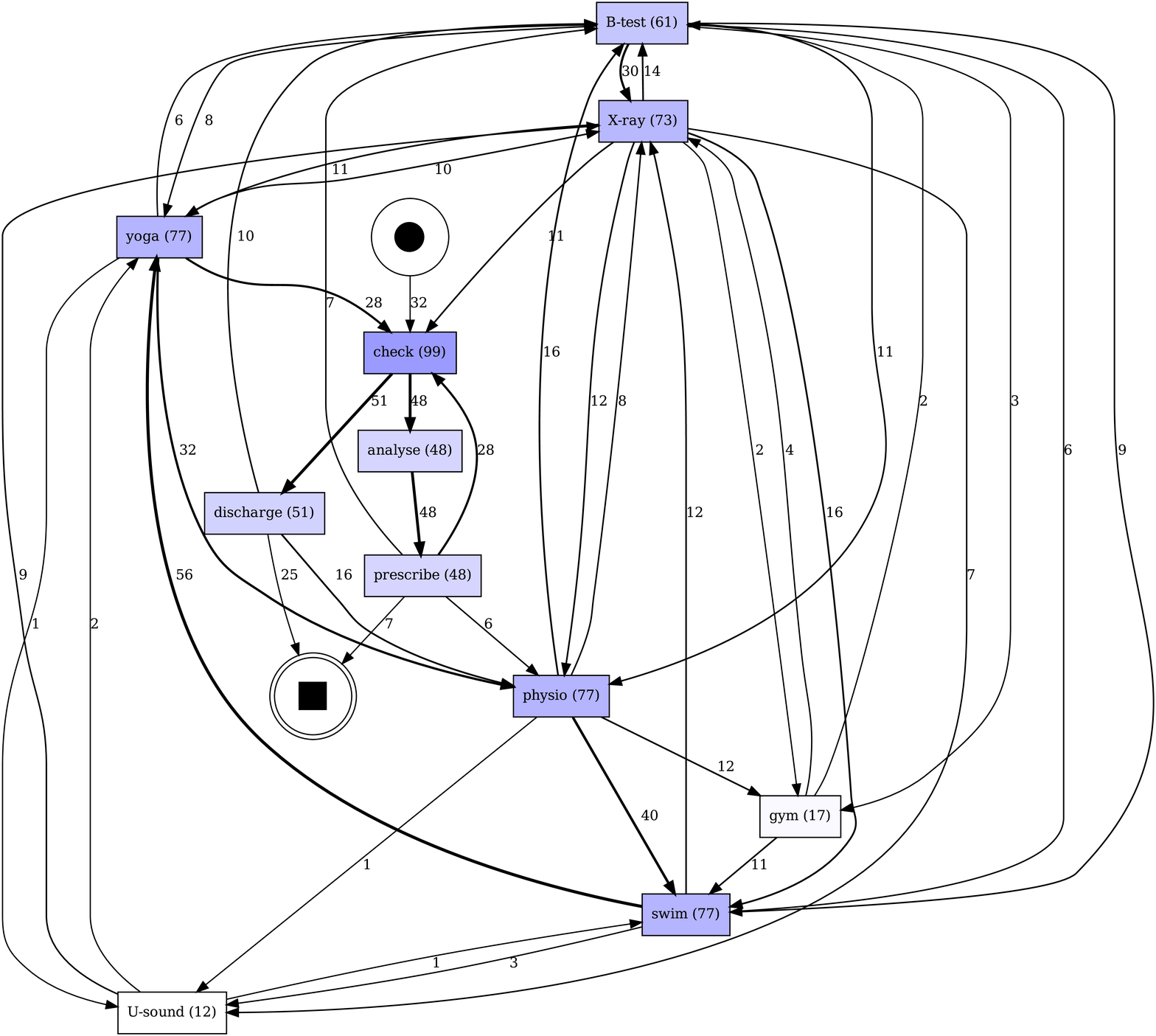}
		\caption{\small{32 traces}}
		\label{fig:dfg_32cases}
	\end{subfigure}
	\vspace{-3mm}
	\caption{\small{DFGs that describe the health surveillance process (readability not intended).}}
	\label{fig:motivating_std_dfgs}
	\vspace{-7mm}
\end{figure}

\Cref{fig:motivating_std_dfgs} shows three directly-follows graphs (DFGs)~\cite{Aalst2016} that describe control flow dependencies between the activities of the health surveillance process discovered from 2, 8, and 32 traces from the log, where a \emph{trace} is a sequence of all events with the same case attribute ordered by their timestamps.
The complexity of the DFGs, defined as the number of nodes and arcs, grows as they represent more data, where the model in \cref{fig:dfg_32cases} is an example spaghetti model.

\begin{figure}[b]
	\vspace{-7mm}
	\centering
	\begin{subfigure}[t]{0.54\textwidth}
		\centering
		\begin{tikzpicture}[scale=0.8, every node/.style={scale=0.8}, node distance=8mm, on grid, thick, >=stealth', bend angle=22, 
			every place/.style={minimum size=5mm, thick, fill=blue!10, draw, drop shadow}, 
			every transition/.style={thick, fill=red!10, draw, minimum size=5mm, drop shadow}, 
			fired transition/.style={transition, fill=green!30}, 
			enabled transition/.style={transition, fill=red!30}, 
			silent transition/.style={transition, fill=black, minimum width=2mm}, 
			every label/.style= {black}]
			\node [silent transition] (t2) [label={[text width=10mm,align=center]above:{$t_2$}}] {};
			\node [place,tokens=0]    (p3) [above right=of t2,label={[text width=10mm,align=center]above:{$p_3$}}] {}
			edge [post]   (t2);
			\node [place,tokens=0]    (p4) [below right=of t2,label={[text width=10mm,align=center]above:{$p_4$}}] {}
			edge [post]   (t2);
			\node [transition]        (t3) [right=of p3,label={[text width=10mm,align=center]above:{$t_3$}}] {\scriptsize{\,\,\textsl{a3}\,\,}}
			edge [post]   (p3);
			\node [transition]        (t4) [right=of p4,label={[text width=10mm,align=center]above:{$t_4$}}] {\scriptsize{\,\,\textsl{a2}\,\,}}
			edge [post]   (p4);
			\node [place,tokens=0]    (p2) [below left=of t2,label={[text width=10mm,align=center]above:{$p_2$}}] {}
			edge [pre]   (t2);
			\node [silent transition] (t1) [above left=of p2,label={[text width=10mm,align=center]above:{$t_1$}}] {}
			edge [post]  (p2);
			\node [place,tokens=1]    (p1) [above left=of t1,label={[text width=10mm,align=center]above:{$p_1$}}] {}
			edge [post]  (t1);
			\node [transition]        (t5) [below=of t4,label={[text width=10mm,align=center]above:{$t_5$}}] {\scriptsize{\,\,\textsl{a1}\,\,}}
			edge [pre, bend left]   (p2);
			\node [place,tokens=0]    (p5) [right=of t3,label={[text width=10mm,align=center]above:{$p_5$}}] {}
			edge [post]   (t3);
			\node [place,tokens=0]    (p6) [right=of t4,label={[text width=10mm,align=center]above:{$p_6$}}] {}
			edge [post]   (t4);
			\node [silent transition] (t6) [below right=of p5,label={[text width=10mm,align=center]above:{$t_6$}}] {}
			edge [post]   (p5)
			edge [post]   (p6);
			\node [place,tokens=0]    (p7) [below right=of t6,label={[text width=10mm,align=center]above:{$p_7$}}] {}
			edge [pre, bend left]   (t5)
			edge [post]   (t6);
			\node [silent transition] (t7) [above right=of p7,label={[text width=10mm,align=center]above:{$t_7$}}] {}
			edge [pre]  (p7);
			\node [place,tokens=0]    (p8) [above right=of t7,label={[text width=10mm,align=center]above:{$p_8$}}] {}
			edge [pre]  (t7);
		\end{tikzpicture}
		\caption{}
		\label{fig:net:system}
	\end{subfigure}
	\begin{subfigure}[t]{0.44\textwidth}
		\centering
		\begin{tikzpicture}[scale=0.8, every node/.style={scale=0.8}, node distance=8mm, on grid, thick, >=stealth', bend angle=22, 
			every place/.style={minimum size=5mm, thick, fill=blue!10, draw, drop shadow}, 
			every transition/.style={thick, fill=red!10, draw, minimum size=5mm, drop shadow}, 
			fired transition/.style={transition, fill=green!30}, 
			enabled transition/.style={transition, fill=red!30}, 
			silent transition/.style={transition, fill=black, minimum width=2mm}, 
			every label/.style= {black}]
			\node [transition]        (t5) [above left=of p6,label={[text width=10mm,align=center]above:{$t_5$}}] {\scriptsize{\,\,\emph{d}\,\,}};
			\node [transition]				(t6) [below=of t5,label={[text width=10mm,align=center]above:{$t_6$}}] {\scriptsize{\,\,\emph{a}\,\,}};
			\node [silent transition] (t7) [below=of t6,label={[text width=10mm,align=center]above:{$t_7$}}] {};
			\node [place,tokens=0]    (p6) [right=of t6,label={[text width=10mm,align=center]above:{$p_6$}}] {}
			edge [pre]   (t5)
			edge [pre]   (t7);
			\node [place,tokens=0]    (p3) [left=of t5,label={[text width=10mm,align=center]above:{$p_3$}}] {}
			edge [post]  (t5)
			edge [post]  (t6);
			\node [place,tokens=0]    (p4) [left=of t6,label={[text width=10mm,align=center]above:{$p_4$}}] {}
			edge [pre]   (t6);
			\node [place,tokens=0]    (p5) [left=of t7,label={[text width=10mm,align=center]above:{$p_5$}}] {}
			edge [post]  (t7);
			\node [transition]				(t2) [left=of p3,label={[text width=10mm,align=center]above:{$t_2$}}] {\scriptsize{\,\,\emph{c}\,\,}}
			edge [post]  (p3);
			\node [transition]				(t3) [left=of p4,label={[text width=10mm,align=center]above:{$t_3$}}] {\scriptsize{\,\,\emph{p}\,\,}}
			edge [pre]   (p4)
			edge [post]  (p5);
			\node [silent transition] (t4) [left=of p5,label={[text width=10mm,align=center]above:{$t_4$}}] {}
			edge [pre]   (p5);
			\node [place,tokens=0]    (p2) [left=of t3,label={[text width=10mm,align=center]above:{$p_2$}}] {}
			edge [post]	 (t2)
			edge [pre]	 (t4);
			\node [silent transition] (t1) [left=of p2,label={[text width=10mm,align=center]above:{$t_1$}}] {}
			edge [post]  (p2);
			\node [place,tokens=1]    (p1) [left=of t1,label={[text width=10mm,align=center]above:{$p_1$}}] {}
			edge [post]  (t1);
		\end{tikzpicture}
		\caption{}
		\label{fig:a1:net:system}
	\end{subfigure}
	\vspace{-2mm}
	\caption{\small{Petri nets that describe: (a) interactions of the three agent types \textsl{a1} (doctor \textsl{d1}), \textsl{a2} (\textsl{d2} and \textsl{d4}), and \textsl{a3} (\textsl{d3} and \textsl{d5}) and (b) agent type \textsl{a1} from the health surveillance process with transition labels check (\emph{c}), analyze (\emph{a}), prescribe (\emph{p}), and discharge (\emph{d}).}}
	\label{fig:AM:running:example}
	\vspace{-2mm}
\end{figure}
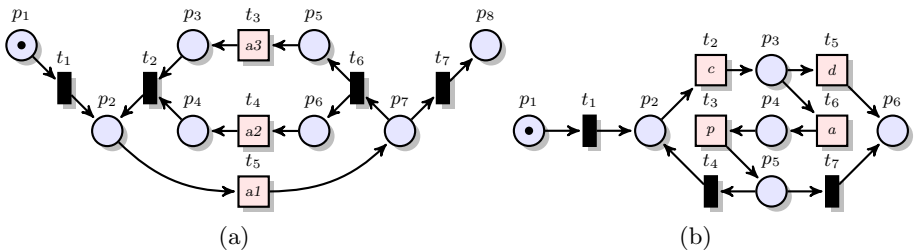

\begin{figure}[t]
	\vspace{-4mm}
	\centering
	\includegraphics[width=1.0\textwidth]{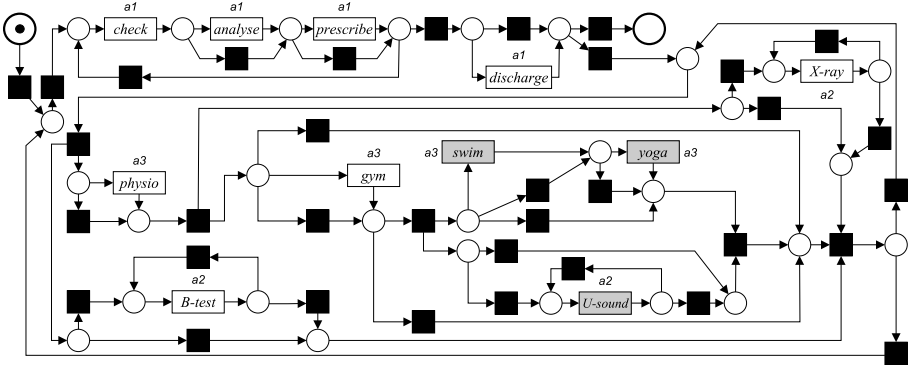}
	\vspace{-7mm}
	\caption{\small{The IM net discovered from 32 traces of the example health surveillance process.}}
	\label{fig:soa:running_example_viz}
	\vspace{-5mm}
\end{figure}

\Cref{fig:net:system} and \cref{fig:a1:net:system} show the interaction net and one agent net that describes one of the three \emph{agent types} discovered by Agent Miner from the events of the 32 traces used to construct the DFG in \cref{fig:dfg_32cases} captured as Petri nets~\cite{Reisig2013}.
Besides providing an alternative, modular perspective on the process, that is, an explicit representation of process participants and their interactions, these models, similar to the DFGs in \cref{fig:motivating_std_dfgs}, describe control flow dependencies between the activities.
These dependencies can be captured explicitly in an integrated Petri net, called the Multi-Agent System (MAS) net, obtained by refining labeled transitions of the interaction net with the corresponding agent nets.
Interestingly, this MAS net is smaller and represents the traces encoded in the event data more faithfully than the Petri net constructed from the same 32 traces using Inductive Miner (IM)~\cite{leemans2013discovering}---a conventional process discovery algorithm.
The MAS net has 155 nodes and arcs, while the IM net shown in \cref{fig:soa:running_example_viz} has 175 nodes and arcs.
The precision of the MAS net and the IM net is 0.37 and 0.16, respectively.
To support model comparison, both algorithms were configured to discover perfectly fitting models (recall of 1.0).
A model has good precision if it does not replay traces not recorded in the log and a good recall if it replays many traces from the log, where the values closer to 1.0 indicate better models.
Precision and recall were measured for the 32 traces used to construct the models using the entropy-based approach~\cite{PolyvyanyyACGKL20,PolyvyanyySWCM20}, the only existing measures that guarantee that better discovered models result in better measurements~\cite{DBLP:journals/topnoc/SyringTA19}.

A MAS net often does not require an increase in size to represent more data well due to an ability of an agent system to simulate the non-decreasing complexity of the behavior of a system~\cite{tour2021agent,Wolfram1984}.
For instance, the same MAS net we discovered from the 32 traces has precision of 0.43 and recall of 1.0 if measured for all 1\,024 traces.
This example confirms that Agent Miner can construct modular, fitting, and precise models of process participants and their interactions.
Finally, while the interaction and agent nets enable dedicated analyses of the respective artifacts, like identification of repetitive work handovers and verification of safeness and liveness of individual agent nets, the information about agents is scattered in the IM net, see the \emph{U-sound} activity by agent \textsl{a2} amidst activities \emph{swim} and \emph{yoga} by agent \textsl{a3} in \cref{fig:soa:running_example_viz}, which hinders the analysis of individual agents and their interactions. 
The input log, models, and measurements discussed in this section are publicly available~\cite{Tour2023data}.

\vspace{-3mm}
\section{Preliminaries}
\label{sec:preliminaries}
\vspace{-2mm}

This section introduces Petri nets (\cref{subsec:petri:nets}) and event logs (\cref{subsec:event_trace_log_trace}). 

\vspace{-3mm}
\subsection{Petri Nets and Workflow Nets}
\label{subsec:petri:nets}
\vspace{-2mm}

Petri nets formalism suits well for describing models of distributed systems~\cite{Reisig2013}.

\vspace{-1mm}
\begin{define}{Petri nets}{def:petri:net}{\quad\\}
	A \emph{(labeled) Petri net}, or a \emph{net}, $N$ is a quintuple $\tuple{P,T,F,\Lambda,\lambda}$, where
	$P$ is a finite set of \emph{places},
	$T$ is a finite set of \emph{transitions},
	$F \subseteq (P \times T) \cup (T \times P)$ is the \emph{flow relation},
	$\Lambda$ is a set of \emph{labels}, such that $\tau \in \Lambda$ is the \emph{silent label} and sets $P$, $T$, and $\Lambda$ are pairwise disjoint, and 
	$\func{\lambda}{T}{\Lambda}$ is the \emph{labeling function}.
\end{define}
\vspace{-1mm}

\noindent
If $\funcCall{\lambda}{t}=\tau$, $t \in T$, $t$ is \emph{silent}; otherwise $t$ is \emph{observable}.
Observable transitions represent activities from the problem domain, and silent transitions encode internal actions of the system. 
A \emph{marking} $M$ of a net encodes its state and is a multiset over its places.
%
\Cref{fig:net:system} shows a Petri net $N=\tuple{P,T,F,\Lambda,\lambda}$, with eight places ($P=\set{p_1,\ldots,p_8}$) and seven transitions ($T=\set{t_1,\ldots,t_7}$).
In the graphical notation, places are drawn as circles, while transitions as squares or rectangles.
Transitions $t_1$, $t_2$, $t_6$, and $t_7$ are silent, shown as black rectangles.
The labeling function assigns labels \textsl{a1}, \textsl{a2}, and \textsl{a3} to transitions $t_5$, $t_4$, and $t_3$, respectively.
The flow relation is shown as directed arcs.
A marking is denoted as an arrangement of black dots, called \emph{tokens}, inside of the corresponding places. 
Marking $M$ shown in \cref{fig:net:system} is the multiset $\mset{p_1}$; see one black dot in place $p_1$.

Let $n \in P \,\cup\, T$ be a place or transition, then by 
$\bullet n \!=\! \setbuilder{x \in (P \cup T)}{(x,n) \in F}$ we denote its \emph{preset} and by 
$n\bullet \!=\! \setbuilder{x \in (P \cup T)}{(n,x) \in F}$ we denote its \emph{postset}. 
Let $N=\tuple{P,T,F,\Lambda,\lambda}$ be a net.
A transition $t \in T$ is \emph{enabled} in a marking $M$ of $N$, denoted by $\enabled{N}{M}{t}$, if every \emph{input place} of $t$ contains at least one token, \ie 
$\predicate{\forall}{p \in \bullet t}{\funcCall{M}{p} > 0}$; by $\funcCall{M}{p}$ we denote the multiplicity  of $p$ in $M$.
An enabled transition $t \in T$ can occur.
An \emph{occurrence} of $t$ leads to a fresh marking $M' = (M \setminus\bullet t) \uplus t\bullet$ of $N$, denoted by $\occurrence{N}{M}{t}{M'}$.

Workflow nets are special nets used for modeling workflows~\cite{Aalst2016}.
A \emph{workflow net} is a Petri net $\tuple{P,T,F,\Lambda,\lambda}$ with 
a dedicated \emph{initial place} $i \in P$, $\bullet i=\emptyset$, 
a dedicated \emph{final place} $f \in P$, $f\bullet=\emptyset$, and
every place and transition on the vertex sequence of some directed walk from $i$ to $f$ in graph $\pair{P \cup T}{F}$.
Marking $\mset{i}$ is the initial marking of a workflow net.
A workflow net is \emph{safe} if every marking reachable from $\mset{i}$ via a sequence of transition occurrences is a set.
It is \emph{sound} if every transition of the net can occur in some sequence of transition occurrences that starts in $\mset{i}$, 
every sequence of transition occurrences that starts in $\mset{i}$ can be extended to ``put'' a token in the final place, and once there is a token in the final place no other places hold tokens~\cite{Aalst2016}.
\Cref{fig:net:system} and \cref{fig:a1:net:system} show safe and sound workflow nets with final markings $\mset{p_8}$ and $\mset{p_6}$, respectively.

\vspace{-3mm}
\subsection{Events, Event Logs, and Traces}
\label{subsec:event_trace_log_trace}
\vspace{-2mm}

\begin{wraptable}{r}{0.44\linewidth}
	\vspace{-13.5mm}
	\caption{\small{An event selection.}}
	\vspace{-3mm}
	\scriptsize
	\centering
	\begin{tabular}{c|c|c|c|c}
		\textbf{event} & \textbf{timestamp} & \textbf{case} & \textbf{activity} & \,\textbf{agent}\, \\
		\hline
		\hline
		$e^a$    & 30 Mar 16:34              & \textsl{case1}         & \emph{check}            & \textsl{d1}/\textsl{a1}         \\
		$e^b$    & 30 Mar 16:35              & \textsl{case1}         & \emph{analyze}          & \textsl{d1}/\textsl{a1}         \\
		$e^c$    & 31 Mar 16:35              & \textsl{case2}         & \emph{check}            & \textsl{d1}/\textsl{a1}         \\
		$e^d$    & 01 Apr 08:22              & \textsl{case2}         & \emph{analyze}          & \textsl{d1}/\textsl{a1}         \\
		$e^e$    & 01 Apr 16:05              & \textsl{case1}         & \emph{prescribe}        & \textsl{d1}/\textsl{a1}         \\
		$e^f$    & 03 Apr 11:55              & \textsl{case1}         & \emph{B-test}           & \textsl{d4}/\textsl{a2}         \\
		$e^g$    & 03 Apr 16:59              & \textsl{case1}         & \emph{X-ray}            & \textsl{d4}/\textsl{a2}         \\
		$e^h$    & 06 Apr 10:02              & \textsl{case1}         & \emph{physio}           & \textsl{d3}/\textsl{a3}         \\
		$e^i$    & 06 Apr 11:01              & \textsl{case1}         & \emph{swim}             & \textsl{d3}/\textsl{a3}         \\
		$e^j$    & 07 Apr 15:55              & \textsl{case1}         & \emph{yoga}             & \textsl{d3}/\textsl{a3}         \\
		$e^k$    & 07 Apr 11:11              & \textsl{case1}         & \emph{physio}           & \textsl{d3}/\textsl{a3}         \\
		$e^l$    & 10 Apr 13:13              & \textsl{case1}         & \emph{swim}             & \textsl{d3}/\textsl{a3}         \\
		$e^m$    & 10 Apr 15:05              & \textsl{case1}         & \emph{yoga}             & \textsl{d3}/\textsl{a3}         \\
		$e^n$    & 11 Apr 09:12              & \textsl{case1}         & \emph{physio}           & \textsl{d3}/\textsl{a3}         \\
		$e^o$    & 11 Apr 10:05              & \textsl{case1}         & \emph{swim}             & \textsl{d3}/\textsl{a3}         \\
		$e^p$    & 13 Apr 11:03              & \textsl{case1}         & \emph{yoga}             & \textsl{d3}/\textsl{a3}         \\
		$e^q$    & 13 Apr 14:57              & \textsl{case1}         & \emph{check}            & \textsl{d1}/\textsl{a1}         \\
		$e^r$    & 16 Apr 12:11              & \textsl{case1}         & \emph{analyze}          & \textsl{d1}/\textsl{a1}         \\
		$e^s$    & 17 Apr 10:03              & \textsl{case1}         & \emph{prescribe}        & \textsl{d1}/\textsl{a1}         \\
		$e^t$    & 17 Apr 16:36              & \textsl{case2}         & \emph{prescribe}        & \textsl{d1}/\textsl{a1}        
	\end{tabular}
	\label{fig:running_example_events}
	\vspace{-11mm}
\end{wraptable}

An \emph{event log}, or \emph{log}, represents real-world processes recorded by an information system.
In process mining, events are often organized into time-ordered sequences, called \emph{traces}.
As explained in~\cref{sec:motivating_example}, in our work, each event has at least four attributes: \emph{timestamp}, \emph{activity}, \emph{case}, and \emph{agent} (either instance or type).
Let $\mathcal{A}$ be the universe of attribute names, with $\set{\mathit{timestamp}, \mathit{activity}, \mathit{case}, \mathit{agent}}\!\!\subseteq\!\!\mathcal{A}$.
Let $\mathcal{V}$ be the universe of attribute values.
Then, an \emph{event} is an attribute function $\func{e}{\mathcal{A}}{\mathcal{V}}$ that maps attribute names to attribute values.
%
By $\mathcal{E}$ we denote the universe of events.
An \emph{event selection} $S \subseteq \mathcal{E}$ is a finite set of events.
Without loss of generality, we assume that events have unique timestamps.
%
\Cref{fig:running_example_events} defines event selection $S=\set{{e^a}, {\ldots}, {e^t}}$.
Each row of the table describes one event.
For example, $e^t = \set{\pair{\mathit{timestamp}}{\textnormal{17 Apr 16:36}}, \pair{\mathit{activity}}{\mathit{prescribe}}, \pair{\mathit{case}}{\textsl{case2}}, \pair{\mathit{agent}}{\textsl{a1}}}$ is the event from the last row of \Cref{fig:running_example_events}; the agent attribute specifies agent type.

Let $X$ be a finite non-empty set. 
A \emph{partition} of $X$ is a set $\Pi$ of disjoint subsets of $X$ such that the union of the subsets equals $X$; the subsets are \emph{parts} of $\Pi$. 
Partition $\Pi$ can be defined by an equivalence relation $\sim\, \subseteq X \times X$ such that if two events $e_i, e_j \in X$ are equivalent under $\sim$, that is, it holds that $\pair{e_i}{e_j} \in\, \sim$, then $e_i$ and $e_j$ belong to the same part of $\Pi$.
We use notation $\Pi = X/\!\sim$ to denote that partition $\Pi$ is defined by the equivalence relation $\sim$ over $X$.
Against this background, we define a trace of an event selection as follows.

\vspace{-1mm}
\begin{define}{Traces induced by partitions}{def:trace}{\quad\\}
	The \emph{trace} $\sigma$ of event selection $S$ induced by part $\pi$ of partition $\Pi = S /\! \sim$ is the ordered by timestamps sequence of all and only events in $\pi$.
\end{define}
\vspace{-1mm}

\noindent
We refer to $\Pi$ and $\sim$ as the \emph{trace partition} and the \emph{trace relation} that induce $\sigma$.
The \emph{trace set} of $S$ induced by $\Pi = S /\!\sim$ is the set of traces $\Sigma$ that for each part $\pi$ of $\Pi$ contains the trace of $S$ induced by $\pi$, and contains no other traces.

\vspace{-1mm}
\begin{define}{Event logs}{def:event_log}{\quad\\}
	An \emph{event log}, or \emph{log}, of event selection $S$ induced by partition $\Pi = S /\! \sim$ is a triple $\tuple{S,\Sigma,\nu}$, where
	$\Sigma \subset S^*$ is the trace set induced by $\Pi$, and
	$\func{\nu}{S}{\mathcal{V}}$ is a naming function that assigns names to events in $S$.
\end{define}
\vspace{-1mm}

\noindent
Multiple event logs induced by different partitions and naming functions can be defined.
For example, in classical process discovery, traces are induced by case attributes.
Let $\mathcal{C}$ be a universe of cases.
The \emph{case trace set} of event selection $S$ is the trace set of $S$ defined by relation $\sim^\textbf{c}\, = \setbuilder{\pair{e_i}{e_j} \in S \times S}{\funcCall{e_i}{\mathit{case}} = \funcCall{e_j}{\mathit{case}}}$, denoted by $\Sigma^\textbf{c}$;
we refer to traces in $\Sigma^\textbf{c}$ as \emph{case traces}.
For instance, the case trace set of the event selection in~\cref{fig:running_example_events} consists of two case traces, namely
$\sequence{e^a,e^b,e^e,e^f,e^g,e^h,e^i,e^j,e^k,e^l,e^m,e^n,e^o,e^p,e^q,e^r,e^s}$ and $\sequence{e^c,e^d,e^t}$.

The naming function of an event log specifies the names of the events used by the discovery algorithms to identify activity names in the constructed process models.
In process discovery, events are often identified by their activity attributes, that is, $\nu=\setbuilder{\pair{e}{\funcCall{e}{\mathit{activity}}}}{e \in S}$.
In general, other naming functions can be used.
We will use this feature in the subsequent sections.

\vspace{-3mm}
\section{Agent Miner}
\label{sec:agent_miner}
\vspace{-2mm}

In this section, we introduce the core notions required to define the Agent Miner algorithm (\cref{subsec:mas_traces_and_logs}) and present the algorithm (\cref{sec:algorithm}).

\vspace{-3mm}
\subsection{Agent Logs and Nets}
\label{subsec:mas_traces_and_logs}
\vspace{-2mm}

A trace of an \emph{agent trace set} is a sequence of events that refer to the same case, are performed by the same agent (identified by the agent attribute), and are not interrupted by an event from the same case performed by a different agent.

\vspace{-1.5mm}
\begin{define}{Agent trace sets}{def:agent-trace}{\quad\\}
	The \emph{agent trace set} of event selection $S$ is the trace set of $S$ induced by the partition of $S$ defined by relation 
	$\sim=\setbuilder{\pair{e_i}{e_j} \in S \times S}{%
		(\funcCall{e_i}{\mathit{agent}}=\funcCall{e_j}{\mathit{agent}})\right.$\newline $\left.\land\;\;
		(\funcCall{e_i}{\mathit{case}}=\funcCall{e_j}{\mathit{case}}) \;\;\land\;\;
		(\neg\exists\, e_k \in S:%
		(%
		(%
		(\funcCall{e_i}{\mathit{timestamp}}<\funcCall{e_k}{\mathit{timestamp}}<\right.$\newline $\left.\funcCall{e_j}{\mathit{timestamp}})%
		\;\;\lor\;\;%
		(\funcCall{e_j}{\mathit{timestamp}}\,\,<\,\,\funcCall{e_k}{\mathit{timestamp}}\,\,<\,\,\funcCall{e_i}{\mathit{timestamp}})%
		)%
		\;\;\land\,\right.$\newline $\left.%
		(%
		\funcCall{e_i}{\mathit{agent}}\neq\funcCall{e_k}{\mathit{agent}}%
		\,\land\,%
		\funcCall{e_i}{\mathit{case}}=\funcCall{e_k}{\mathit{case}}%
		)%
		)%
		)%
	}.%
	$%
\end{define}
\vspace{-1mm}

\noindent
The set 
$\{\sequence{e^a,e^b,e^e},\sequence{e^c,e^d,e^t},$ $\sequence{e^f,e^g},\sequence{e^h,e^i,e^j,e^k,e^l,e^m,e^n,e^o,e^p},\sequence{e^q,e^r,e^s}\}$
is the agent trace set of the event selection in \Cref{fig:running_example_events}.
For instance, in trace $\sequence{e^a,e^b,e^e}$, all events are from \textsl{case1}, performed by agent \textsl{a1}, and, though interrupted by events $e^c$ and $e^d$, the latter events are from a different case.
Note that, by definition, relation $\sim$ from \Cref{def:agent-trace} is an equivalence relation.

Next, we define several useful logs.
Traces of an interaction log are composed of events that allow identifying all handovers of work between agents.

\begin{figure}[b]
	\vspace{-4mm}
	\begin{center}
		\begin{tikzpicture}[scale=0.8, every node/.style={scale=0.8}, node distance=9mm, on grid, thick, >=stealth', bend angle=22, 
			every place/.style={minimum size=5mm, thick, fill=blue!10, draw, drop shadow}, 
			every transition/.style={thick, fill=red!10, draw, minimum size=5mm, drop shadow}, 
			fired transition/.style={transition, fill=green!30}, 
			enabled transition/.style={transition, fill=red!30}, 
			silent transition/.style={transition, fill=black, minimum width=2mm}, 
			every label/.style= {black}]
			\node [transition] (t2) [label={[text width=10mm,align=center]above:{$t_2$}}] {\scriptsize{\,\,\textsl{a3}\,\,}};
			\node [place,tokens=0]    (p2) [left=of t2,label={[text width=10mm,align=center]above:{$p_2$}}] {}
			edge [pre]   (t2);
			\node [silent transition] (t1) [left=of p2,label={[text width=10mm,align=center]above:{$t_1$}}] {}
			edge [post]  (p2);
			\node [place,tokens=1]    (p1) [left=of t1,label={[text width=10mm,align=center]above:{$p_1$}}] {}
			edge [post]  (t1);
			\node [place,tokens=0]    (p3) [right=of t2,label={[text width=10mm,align=center]above:{$p_3$}}] {}
			edge [post]   (t2);
			\node [transition] (t3) [right=of p3,label={[text width=10mm,align=center]above:{$t_3$}}] {\scriptsize{\,\,\textsl{a2}\,\,}}
			edge [post]   (p3);
			\node [place,tokens=0]    (p4) [right=of t3,label={[text width=10mm,align=center]above:{$p_4$}}] {}
			edge [post]   (t3);
			\node [silent transition] (t4) [right=of p4,label={[text width=10mm,align=center]above:{$t_4$}}] {}
			edge [pre]  (p4);
			\node [place,tokens=0]    (p5) [right=of t4,label={[text width=10mm,align=center]above:{$p_5$}}] {}
			edge [pre]  (t4);
			\node [transition]        (t5) [below=of p3,label={[text width=10mm,align=center]above:{$t_5$}}] {\scriptsize{\,\,\textsl{a1}\,\,}}
			edge [pre, bend left]   (p2)
			edge [post, bend right]   (p4);
		\end{tikzpicture}
		\vspace{-2mm}
		\caption{\small{An i-net.}}
		\label{fig:i_net}
		\vspace{-2mm}
	\end{center}
\end{figure}
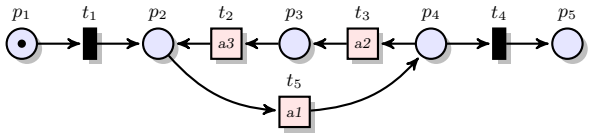

\vspace{-1.5mm}
\begin{define}{Interaction logs}{def:interaction_log}{\quad\\}
	The \emph{interaction log} of event selection $S$ is the triple $\tuple{\bar{S},\Sigma,\nu}$, where
	$\bar{S}$ is the \emph{agent event selection} composed of the first events of all the traces in the agent trace set $\Delta$ of $S$, that is,
	$\bar{S} = \setbuilder{\funcCall{\delta}{1}}{\delta \in \Delta}$,
	$\Sigma$ is the case trace set of $\bar{S}$, and
	$\nu = \setbuilder{\pair{e}{\funcCall{e}{\mathit{agent}}}}{e \in \bar{S}}$ names each event by the corresponding agent.
\end{define}
\vspace{-1.5mm}

\noindent
The interaction log of the event selection from \Cref{fig:running_example_events} is, therefore, the event log $\tuple{\bar{S},\Sigma,\nu}$, where
$\bar{S} = \set{e^a,e^c,e^f,e^h,e^q}$, 
$\Sigma = \set{\sequence{e^a,e^f,e^h,e^q}, \sequence{e^c}}$, and 
$\nu = \set{\pair{e^a}{\textsl{a1}}, \pair{e^c}{\textsl{a1}}, \pair{e^f}{\textsl{a2}}, \pair{e^h}{\textsl{a3}}, \pair{e^q}{\textsl{a1}}}$.
For instance, trace $\sequence{e^a,e^f,e^h,e^q}$ in $\Sigma$ suggests that agent $\textsl{a1}$ starts \textsl{case1}; note that $\funcCall{e^a}{agent}=\textsl{a1}$ and $\funcCall{e^a}{case}=\textsl{case1}$.
Then, agent $\textsl{a2}$ takes over the work on \textsl{case1}.
Agent $\textsl{a2}$ then passes the work on \textsl{case1} to agent \textsl{a3}, who later hands work back to agent \textsl{a1}.

Traces in the agent trace set done by the same agent compose its agent log.

\vspace{-1.5mm}
\begin{define}{Agent logs}{def:agent_log}{\quad\\}
	The \emph{agent log} of agent $\textsl{a}$ and event selection $S$ is the triple $\tuple{S^\textsl{a},\Sigma,\nu}$, where
	$S^\textsl{a}=\setbuilder{e \in S}{\funcCall{e}{\mathit{agent}}=\textsl{a}}$,
	$\Sigma$ is the set of traces in the agent trace set $\Delta$ of $S$ performed by $\textsl{a}$, that is, 
	$\Sigma=\setbuilder{\delta\in\Delta}{\forall\, i \in \intintervalcc{1}{\seqLength{\delta}} : \funcCall{\funcCall{\delta}{i}}{\mathit{agent}}=\textsl{a}}$, and 
	$\nu = \setbuilder{\pair{e}{\pair{\textsl{a}}{\funcCall{e}{\mathit{activity}}}}}{e \in S^\textsl{a}}$ is the naming function that names an event by the pair comprising its agent and activity attributes.
\end{define}
\vspace{-1.5mm}

\noindent
The agent log of agent $\textsl{a1}$ and the event selection from \Cref{fig:running_example_events} is defined by the triple $\tuple{S^\textsl{a1},\Sigma,\nu}$, where
$S^\textsl{a1} = \set{e^a, e^b, e^c, e^d, e^e, e^q, e^r, e^s, e^t}$,  
$\Sigma = \{{\sequence{e^a,e^b,e^e}},\\ {\sequence{e^c,e^d,e^t}}, {\sequence{e^q,e^r,e^s}}\}$, and
$\nu$ maps events in $S^\textsl{a1}$ to their names and contains, for instance,  
it holds that $\set{\pair{e^a}{\pair{\textsl{a1}}{\mathit{check}}},\pair{e^d}{\pair{\textsl{a1}}{\mathit{analyze}}}} \subset \nu$.

Next, we discuss several classes of workflow nets used by Agent Miner.
An interaction net describes the structure of interactions between agents.

\vspace{-1.5mm}
\begin{define}{Interaction nets}{def:interaction_net}{\quad\\}
	An \emph{interaction net}, or an \emph{i-net}, of event selection $S$ is a workflow net $\tuple{P,T,F,\Lambda,\lambda}$, where 
	$\Lambda = \setbuilder{\textsl{a} \in \mathcal{V}}{\exists\, e \in S : \funcCall{e}{agent}=\textsl{a}} \cup \set{\tau}$. 
\end{define}
\vspace{-1.5mm}

\noindent
\Cref{fig:net:system,fig:i_net} show two i-nets of the event selection from \Cref{fig:running_example_events}. 
They describe alternative ways the agents could have interacted to generate the event data. 
For example, the i-net in \Cref{fig:i_net} suggests that agent $\textsl{a1}$ starts the interaction, and then any number of sequences of interactions of $\textsl{a1}$ with agent $\textsl{a2}$, then of $\textsl{a2}$ with agent $\textsl{a3}$, and finally of $\textsl{a3}$ again with agent $\textsl{a1}$ can occur.

A MAS net describes how agents perform activities in a collaborative setting.

\vspace{-1.5mm}
\begin{define}{Multi-Agent System nets}{def:mas_net}{\quad\\}
	A~\emph{Multi-Agent System (MAS) net} of event selection $S$ is a workflow net $(P,T,F,\\\Lambda,\lambda)$, with 
	$\Lambda = \setbuilder{(\textsl{a},b) \!\in\! \mathcal{V}\!\times\!\mathcal{V}}{\exists\, e \!\in\! S \!:\!\funcCall{e}{\mathit{agent}}\!=\!\textsl{a} \land \funcCall{e}{\mathit{activity}}\!=\!b} \cup \set{\tau}$.
\end{define}
\vspace{-1.5mm}

\noindent
A MAS net of events of a single agent is an agent net of this agent.

\vspace{-1.5mm}
\begin{define}{Agent nets}{def:agent_net}{\quad\\}
	A MAS net of event selection $S$ such that the agent attribute of every event in $S$ is equal to $\textsl{a}$, that is, $\forall\, e\in S : \funcCall{e}{\mathit{agent}}=\textsl{a}$, is an \emph{agent net} of $\textsl{a}$.
\end{define}
\vspace{-1.5mm}

\vspace{-6mm}
\subsection{Algorithm}
\label{sec:algorithm}
\vspace{-1mm}

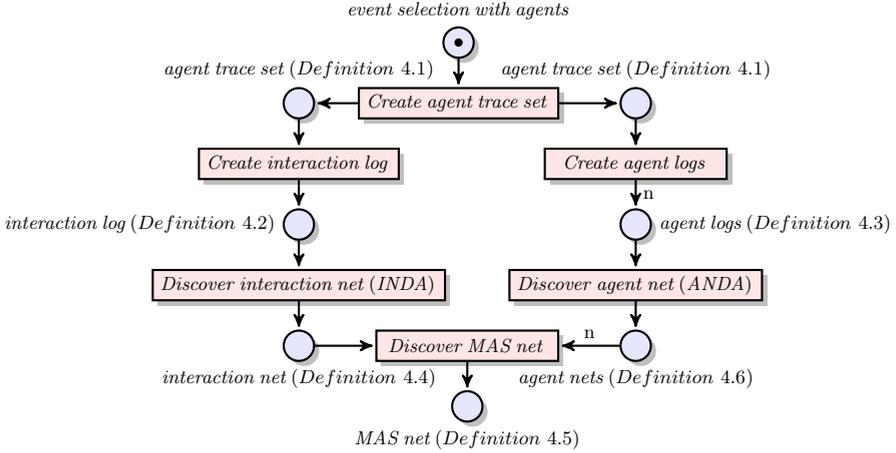
\begin{figure}[t]
	\begin{center}
		\vspace{-2mm}
		\begin{tikzpicture}[scale=0.8, every node/.style={scale=0.8}, node distance=8mm, on grid, thick, >=stealth', bend angle=22, 
			every place/.style={minimum size=5mm, thick, fill=blue!10, draw, drop shadow}, 
			every transition/.style={thick, fill=red!10, draw, minimum width=30mm, minimum height=5mm, drop shadow}, 
			fired transition/.style={transition, fill=green!30}, 
			enabled transition/.style={transition, fill=red!30}, 
			silent transition/.style={transition, fill=black, minimum width=2mm}, 
			every label/.style= {black}]
			\node [place,tokens=0]	(p_mm) [label=below:$\mathit{MAS\,net\,(\cref{def:mas_net})}$] {};
			\node [transition] 		(t_dmm) [above=of p_mm] {
				$\mspace{1mu}$
				$\mathit{Discover\,MAS\,net}$
				$\mspace{3mu}$
			} edge [post]  (p_mm);
			\node [place,tokens=0]	(p_cm1) [right=of t_dmm.east,label=below:$\mathit{agent\,nets\,(\cref{def:agent_net})}$] {} edge [post] node [above] {n} (t_dmm);
			\node [place,tokens=0]	(p_am1) [left=of t_dmm.west,label=below:$\mathit{interaction\,net\,(\cref{def:interaction_net})}$] {} edge [post] (t_dmm);
			\node [transition]		(t_dcm) [above=of p_cm1] {
				$\mspace{1mu}$
				$\mathit{Discover\,agent\,net\,(ANDA)}$
				$\mspace{3mu}$
			} edge [post]  (p_cm1);
			\node [transition]		(t_dam) [above=of p_am1] {
				$\mspace{1mu}$
				$\mathit{Discover\,interaction\,net\,(INDA)}$
				$\mspace{3mu}$
			} edge [post]  (p_am1);
			\node [place,tokens=0]	(p_cl) [above=of t_dcm,label=right:$\mathit{agent\,logs\,(\cref{def:agent_log})}$] {} edge [post]  (t_dcm);
			\node [place,tokens=0]	(p_al) [above=of t_dam,label=left:$\mathit{interaction\,log\,(\cref{def:interaction_log})}$] {} edge [post] (t_dam);
			\node [transition]		(t_ccl) [above=of p_cl] {
				$\mspace{1mu}$
				$\mathit{Create\,agent\,logs}$
				$\mspace{3mu}$
			} edge [post] node [right] {n}  (p_cl);
			\node [transition]	(t_cal) [above=of p_al] {
				$\mspace{1mu}$
				$\mathit{Create\,interaction\,log}$
				$\mspace{3mu}$
			} edge [post] (p_al);
			\node [place,tokens=0]	(p_fc) [above=of t_ccl,label=above:$\mathit{agent\,trace\,set\,(\cref{def:agent-trace})}$] {} edge [post] (t_ccl);
			\node [place,tokens=0]	(p_fa) [above=of t_cal,label=above:$\mathit{agent\,trace\,set\,(\cref{def:agent-trace})}$] {} edge [post] (t_cal);
			\node [transition]		(t_if) [left=of p_fc.west] {
				$\mspace{1mu}$
				$\mathit{Create\,agent\,trace\,set}$
				$\mspace{3mu}$
			} edge [post]  (p_fa) edge [post]  (p_fc);
			\node [place,tokens=1]	(p_mel) [above=of t_if,label=above:$\mathit{event\,selection\,with\,agents}$] {} edge [post]  (t_if);
		\end{tikzpicture}
		\vspace{-3mm}
		\caption{\small{The Agent Miner algorithm.}}
		\vspace{-10mm}
		\label{fig:asm_discovery_overview}
	\end{center}
\end{figure}

\Cref{fig:asm_discovery_overview} defines the Agent Miner algorithm as a workflow net.
It is parameterized by two conventional control flow discovery algorithms: an Agent Net Discovery Algorithm (ANDA) and an Interaction Net Discovery Algorithm (INDA).
Agent Miner takes an event selection as input and produces an interaction net, agent nets, and a MAS net. 
Similar to standard process discovery algorithms, the latter explains how the discovered interaction and agent nets represent the traces identified by the case attribute.
The algorithm has six steps detailed below.

\smallskip
\noindent
\textbf{Create agent trace set.} 
The agent trace set of the input event selection is created by associating each event with an agent trace, see \cref{def:agent-trace}.

\smallskip
\noindent
\textbf{Create interaction log.}
The interaction log is created by selecting the first events of all traces in the agent trace set and constructing the case trace set of this event selection, see \cref{def:interaction_log}.

\smallskip
\noindent
\textbf{Discover interaction net.} 
In this step, INDA is used to discover an i-net, see \cref{def:interaction_net}, from the interaction log.
For example, the i-net in \cref{fig:i_net} was discovered from the interaction log of the event selection in \Cref{fig:running_example_events}.

\smallskip
\noindent
\textbf{Create agent logs.} 
Multiple ($n$) agent logs are created, one for each agent encountered in the input event selection, by extracting the corresponding agent traces from the agent trace set, see \cref{def:agent_log}.

\begin{figure}[t]
	\centering
	\vspace{-2mm}
	\begin{subfigure}[b]{0.8\textwidth}
		\centering
		\begin{tikzpicture}[scale=0.8, every node/.style={scale=0.8}, node distance=14mm, on grid, thick, >=stealth', bend angle=22, 
			every place/.style={minimum size=5mm, thick, fill=blue!10, draw, drop shadow}, 
			every transition/.style={thick, fill=red!10, draw, minimum width=15mm, minimum height=5mm, drop shadow}, 
			fired transition/.style={transition, fill=green!30}, 
			enabled transition/.style={transition, fill=red!30}, 
			silent transition/.style={transition, fill=black, minimum width=2mm}, 
			every label/.style= {black}]
			\node [place,tokens=1] (p1) [label={[text width=10mm,align=center]above:{$p_1$}}] {};
			\node [transition] (t1) [right=of p1,label={[text width=10mm,align=center]above:{$t_1$}}] {\scriptsize{\,(\textsl{a1},$\mathit{check}$)\,}} edge [pre] (p1);
			\node [place,tokens=0] (p2) [right=of t1,label={[text width=10mm,align=center]above:{$p_2$}}] {} edge [pre] (t1);
			\node [transition] (t2) [right=of p2,label={[text width=10mm,align=center]above:{$t_2$}}] {\scriptsize{\,(\textsl{a1},$\mathit{analyze}$)\,}} edge [pre] (p2);
			\node [place,tokens=0] (p3) [right=of t2,label={[text width=10mm,align=center]above:{$p_3$}}] {} edge [pre] (t2);
			\node [transition] (t3) [right=of p3,label={[text width=10mm,align=center]above:{$t_3$}}] {\scriptsize{\,(\textsl{a1},$\mathit{prescribe}$)\,}} edge [pre] (p3);
			\node [place,tokens=0] (p4) [right=of t3,label={[text width=10mm,align=center]above:{$p_4$}}] {} edge [pre]  (t3);
		\end{tikzpicture}
		\caption{\small{Agent net \textsl{a1}}}
		\label{fig:agent_nets_a1}
	\end{subfigure}
	\begin{subfigure}[b]{0.55\textwidth}
		\centering
		\begin{tikzpicture}[scale=0.8, every node/.style={scale=0.8}, node distance=12mm, on grid, thick, >=stealth', bend angle=22, 
			every place/.style={minimum size=5mm, thick, fill=blue!10, draw, drop shadow}, 
			every transition/.style={thick, fill=red!10, draw, minimum width=15mm, minimum height=5mm, drop shadow}, 
			fired transition/.style={transition, fill=green!30}, 
			enabled transition/.style={transition, fill=red!30}, 
			silent transition/.style={transition, fill=black, minimum width=2mm}, 
			every label/.style= {black}]
			\node [place,tokens=1] (p1) [label={[text width=10mm,align=center]above:{$p_1$}}] {};
			\node [transition] (t1) [right=of p1,label={[text width=10mm,align=center]above:{$t_1$}}] {\scriptsize{\,(\textsl{a2},$\mathit{B}$-$\mathit{test}$)\,}} edge [pre] (p1);
			\node [place,tokens=0] (p2) [right=of t1,label={[text width=10mm,align=center]above:{$p_2$}}] {} edge [pre] (t1);
			\node [transition] (t2) [right=of p2,label={[text width=10mm,align=center]above:{$t_2$}}] {\scriptsize{\,(\textsl{a2},$\mathit{X}$-$\mathit{ray}$)\,}} edge [pre] (p2);
			\node [place,tokens=0] (p3) [right=of t2,label={[text width=10mm,align=center]above:{$p_3$}}] {} edge [pre] (t2);
		\end{tikzpicture}
		\caption{\small{Agent net \textsl{a2}}}
		\label{fig:agent_nets_a2}
	\end{subfigure}
	\begin{subfigure}[b]{0.99\textwidth}
		\centering
		\begin{tikzpicture}[scale=0.8, every node/.style={scale=0.8}, node distance=12mm, on grid, thick, >=stealth', bend angle=22, 
			every place/.style={minimum size=5mm, thick, fill=blue!10, draw, drop shadow}, 
			every transition/.style={thick, fill=red!10, draw, minimum width=15mm, minimum height=5mm, drop shadow}, 
			fired transition/.style={transition, fill=green!30}, 
			enabled transition/.style={transition, fill=red!30}, 
			silent transition/.style={transition, fill=black, minimum width=2mm}, 
			every label/.style= {black}]
			\node [place,tokens=1] (p0) [label={[text width=10mm,align=center]above:{$p_1$}}] {};
			\node [silent transition] (t0) [right=of p1,label={[text width=10mm,align=center]above:{$t_1$}}] {} edge [pre] (p0);
			\node [place,tokens=0] (p1) [right=of t0,label={[text width=10mm,align=center]above:{$p_2$}}] {} edge [pre] (t0);
			\node [transition] (t1) [above right=of p1,label={[text width=10mm,align=center]above:{$t_2$}}] {\scriptsize{\,(\textsl{a3},$\mathit{physio}$)\,}} edge [pre] (p1);
			\node [place,tokens=0] (p2) [right=of t1,label={[text width=10mm,align=center]above:{$p_3$}}] {} edge [pre] (t1);
			\node [transition] (t2) [right=of p2,label={[text width=10mm,align=center]above:{$t_3$}}] {\scriptsize{\,(\textsl{a3},$\mathit{swim}$)\,}} edge [pre] (p2);
			\node [place,tokens=0] (p3) [right=of t2,label={[text width=10mm,align=center]above:{$p_4$}}] {} edge [pre] (t2);
			\node [transition] (t3) [right=of p3,label={[text width=10mm,align=center]above:{$t_4$}}] {\scriptsize{\,(\textsl{a3},$\mathit{yoga}$)\,}} edge [pre] (p3);
			\node [place,tokens=0] (p4) [below right=of t3,label={[text width=10mm,align=center]above:{$p_5$}}] {} edge [pre]  (t3);
			\node [silent transition] (t4) [right=of p4,label={[text width=10mm,align=center]above:{$t_5$}}] {} edge [pre] (p4);
			\node [place,tokens=0] (p5) [right=of t4,label={[text width=10mm,align=center]above:{$p_6$}}] {} edge [pre]  (t4);
			\node [silent transition] (t5) [below right=of p2,label={[text width=10mm,align=center]above:{$t_6$}}] {} edge [pre] (p4) edge [post] (p1);
		\end{tikzpicture}
		\caption{\small{Agent net \textsl{a3}}}
		\label{fig:agent_nets_a3}
	\end{subfigure}
	\vspace{-2mm}
	\caption{\small{Three agent nets.}}
	\label{fig:agent_nets}
\end{figure}
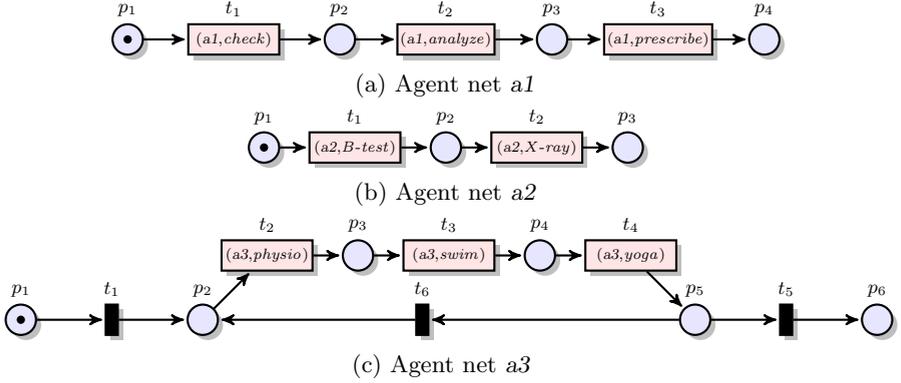

\smallskip
\noindent
\textbf{Discover agent net.} 
In this step, ANDA is used to discover $n$ agent nets (one net is discovered from each agent log), see \cref{def:agent_net}.
\Cref{fig:agent_nets} shows three agent nets discovered from the agent logs of the event selection in \Cref{fig:running_example_events}.

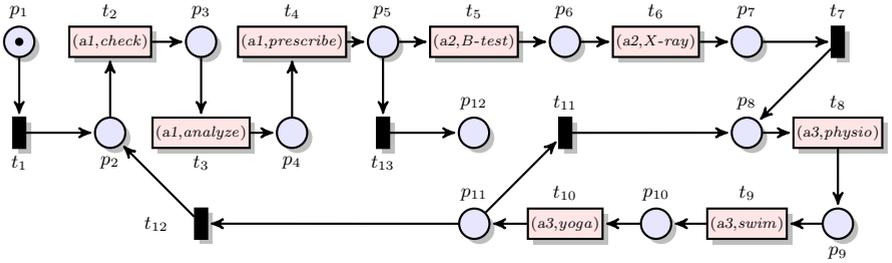
\begin{figure}[t]
	\vspace{-4mm}
	\centering
	\begin{tikzpicture}[scale=0.8, every node/.style={scale=0.8}, node distance=12mm, on grid, thick, >=stealth', bend angle=22, 
		every place/.style={minimum size=5mm, thick, fill=blue!10, draw, drop shadow}, 
		every transition/.style={thick, fill=red!10, draw, minimum width=10mm, minimum height=5mm, drop shadow}, 
		fired transition/.style={transition, fill=green!30}, 
		enabled transition/.style={transition, fill=red!30}, 
		silent transition/.style={transition, fill=black, minimum width=2mm}, 
		every label/.style= {black}]
		\node [place,tokens=1] (p0) [label={[text width=10mm,align=center]above:{$p_1$}}] {};
		\node [silent transition] (t0) [below=of p0,label={[text width=10mm,align=center]below:{$t_1$}}] {} edge [pre] (p0);
		\node [place,tokens=0] (p1) [right=of t0,label={[text width=10mm,align=center]below:{$p_2$}}] {} edge [pre] (t0);
		\node [transition] (t1) [above=of p1,label={[text width=10mm,align=center]above:{$t_2$}}] {\scriptsize{\,(\textsl{a1},$\mathit{check}$)\,}} edge [pre] (p1);
		\node [place,tokens=0] (p2) [right=of t1,label={[text width=10mm,align=center]above:{$p_3$}}] {} edge [pre] (t1);
		\node [transition] (t2) [below=of p2,label={[text width=10mm,align=center]below:{$t_3$}}] {\scriptsize{\,(\textsl{a1},$\mathit{analyze}$)\,}} edge [pre] (p2);
		\node [place,tokens=0] (p3) [right=of t2,label={[text width=10mm,align=center]below:{$p_4$}}] {} edge [pre] (t2);
		\node [transition] (t3) [above=of p3,label={[text width=10mm,align=center]above:{$t_4$}}] {\scriptsize{\,(\textsl{a1},$\mathit{prescribe}$)\,}} edge [pre] (p3);
		\node [place,tokens=0] (p4) [right=of t3,label={[text width=10mm,align=center]above:{$p_5$}}] {} edge [pre]  (t3);
		
		\node [transition] (t4) [right=of p4,label={[text width=10mm,align=center]above:{$t_5$}}] {\scriptsize{\,(\textsl{a2},$\mathit{B}$-$\mathit{test}$)\,}} edge [pre] (p4);
		\node [place,tokens=0] (p5) [right=of t4,label={[text width=10mm,align=center]above:{$p_6$}}] {} edge [pre] (t4);
		\node [transition] (t5) [right=of p5,label={[text width=10mm,align=center]above:{$t_6$}}] {\scriptsize{\,(\textsl{a2},$\mathit{X}$-$\mathit{ray}$)\,}} edge [pre] (p5);
		\node [place,tokens=0] (p6) [right=of t5,label={[text width=10mm,align=center]above:{$p_7$}}] {} edge [pre] (t5);
		\node [silent transition] (t6) [right=of p6,label={[text width=10mm,align=center]above:{$t_7$}}] {} edge [pre] (p6);
		
		\node [place,tokens=0] (p7) [below=of p6,label={[text width=10mm,align=center]above:{$p_8$}}] {} edge [pre] (t6);
		\node [transition] (t7) [right=of p7,label={[text width=10mm,align=center]above:{$t_8$}}] {\scriptsize{\,(\textsl{a3},$\mathit{physio}$)\,}} edge [pre] (p7);
		\node [place,tokens=0] (p8) [below=of t7,label={[text width=10mm,align=center]below:{$p_9$}}] {} edge [pre] (t7);
		\node [transition] (t8) [left=of p8,label={[text width=10mm,align=center]above:{$t_9$}}] {\scriptsize{\,(\textsl{a3},$\mathit{swim}$)\,}} edge [pre] (p8);
		\node [place,tokens=0] (p9) [left=of t8,label={[text width=10mm,align=center]above:{$p_{10}$}}] {} edge [pre] (t8);
		\node [transition] (t9) [left=of p9,label={[text width=10mm,align=center]above:{$t_{10}$}}] {\scriptsize{\,(\textsl{a3},$\mathit{yoga}$)\,}} edge [pre] (p9);
		\node [place,tokens=0] (p10) [left=of t9,label={[text width=10mm,align=center]above:{$p_{11}$}}] {} edge [pre]  (t9);
		\node [silent transition] (t10) [above=of t9,label={[text width=10mm,align=center]above:{$t_{11}$}}] {} edge [pre] (p10) edge [post] (p7);
		\node [silent transition] (t11) [below=of t2,label={[text width=10mm,align=center]left:{$t_{12}$}}] {} edge [pre] (p10) edge [post] (p1);
		
		\node [silent transition] (t12) [below=of p4,label={[text width=10mm,align=center]below:{$t_{13}$}}] {} edge [pre] (p4);
		\node [place,tokens=0] (p11) [right=of t12,label={[text width=10mm,align=center]above:{$p_{12}$}}] {} edge [pre] (t12);
		
	\end{tikzpicture}
	\vspace{-4mm}
	\caption{\small{A MAS net.}}
	\vspace{-7mm}
	\label{fig:mas_net}
\end{figure}

\smallskip
\noindent
\textbf{Discover MAS net.}
Finally, a MAS net, see \cref{def:mas_net}, is constructed by ``embedding'' the agent nets into the i-net and applying the Fusion of Series Places reduction~\cite{Murata1989} to the refined i-net.
\cref{fig:mas_net} shows the embedding of the agent nets in \cref{fig:agent_nets} into the i-net in \cref{fig:i_net}.
The embedding is performed by refining each observable transition in the i-net with the corresponding agent net.
The resulting MAS net describes how agents (doctors) interact to support the execution of cases (treatment of patients) in the health surveillance process introduced in \cref{sec:motivating_example} discovered based on the event selection from \cref{fig:running_example_events}.

Thus, Agent Miner is a divide-and-conquer algorithm that ``divides'' the input event selection into interaction and agent logs and ``conquers'' these logs using ANDA and INDA.
In this work, as ANDA, we used DFG translation to Petri nets (DFG-PN)~\cite{Leemans2019}, while as INDA, we used Inductive Miner (IM)~\cite{leemans2013discovering} and Split Miner (SM)~\cite{Augusto2019} and removed iterations of observable transitions in the obtained i-nets.
This approach follows the agent system paradigm, where agents are sequential machines, and parallelism emerges through collaborations of agents.
If the algorithms guarantee that the constructed models are safe and sound workflow nets, which holds for DFG-PN and IM, every constructed MAS net is guaranteed to be a safe and sound workflow net (cf. Theorem~2 in \cite{DBLP:journals/acta/PolyvyanyyWW11}).

\vspace{-3mm}
\section{Evaluation}
\label{sec:eval}
\vspace{-2mm}

The Agent Miner algorithm (\Cref{fig:asm_discovery_overview}) and its evaluation pipeline (\Cref{fig:eval_pipeline}), including the code, the discovered models, and detailed results, are publicly available~\cite{Tour2023data}.
This section presents and discusses the design (\cref{sec:eval_pipeline}) and results (\cref{sec:eval_results_synthetic}) of an evaluation of Agent Miner using real-world datasets.

\vspace{-3mm}
\subsection{Design}
\label{sec:eval_pipeline}
\vspace{-2mm}

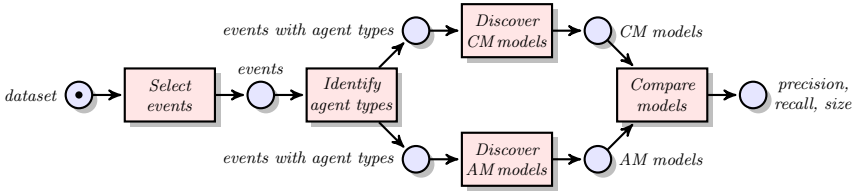
\begin{figure}[h!]
	\vspace{-6mm}
	\begin{center}
		\begin{tikzpicture}[scale=0.7, every node/.style={scale=0.7}, node distance=12mm, on grid, thick, >=stealth', bend angle=22, 
			every place/.style={minimum size=5mm, thick, fill=blue!10, draw, drop shadow}, 
			every transition/.style={thick, fill=red!10, draw, minimum size=5mm, drop shadow}, 
			fired transition/.style={transition, fill=green!30}, 
			enabled transition/.style={transition, fill=red!30}, 
			silent transition/.style={transition, fill=black, minimum width=2mm}, 
			every label/.style= {black}]
			\node [place,tokens=1]    (p1) [label={[text width=13mm,align=right]left:{\emph{dataset}}}] {};
			\node [transition]        (t1) [text centered,right=of p1,text width=1.7cm, minimum height=1cm] {\emph{Select events}} edge [pre]  (p1);
			\node [place,tokens=0]    (p2) [right=of t1,label={[text width=20mm,align=center]above:{\emph{events}}}] {} edge [pre]  (t1);
			\node [transition]        (t2) [text centered,right=of p2,text width=1.7cm, minimum height=1cm] {\emph{Identify agent types}} edge [pre] (p2);
			\node [place,tokens=0]    (p3) [below right=of t2,label={[text width=70mm,align=right]left:{\emph{events with agent types}}}] {} edge [pre] (t2);
			\node [place,tokens=0]    (p4) [above right=of t2,label={[text width=70mm,align=right]left:{\emph{events with agent types}}}] {} edge [pre]  (t2);
			\node [transition]        (t5) [text centered,right=of p3,text width=1.7cm, minimum height=1cm] {\emph{Discover AM\,models}}
			edge [pre]  (p3);
			\node [place,tokens=0]    (p7) [right=of t5,label={[text width=70mm,align=left]right:{\emph{AM models}}}] {}
			edge [pre]  (t5);
			\node [transition]        (t6) [text centered,above right=of p7,text width=1.7cm, minimum height=1cm] {\emph{Compare models}}
			edge [pre]  (p7);
			\node [place,tokens=0]    (p8) [right=of t6,label={[text width=15mm,align=center]right:{\emph{precision, recall, size}}}] {}
			edge [pre]  (t6);
			\node [place,tokens=0]    (p6) [above left =of t6,label={[text width=70mm,align=left]right:{\emph{CM models}}}] {}
			edge [post]  (t6);
			\node [transition]        (t4) [text centered,right=of p4,text width=1.7cm, minimum height=1cm] {\emph{Discover CM\,models}} edge [pre]  (p4) edge [post]  (p6);
			
		\end{tikzpicture}
		\vspace{-6mm}
		\caption{\small{Evaluation pipeline.}}
		\label{fig:eval_pipeline}
		\vspace{-11mm}
	\end{center}
\end{figure}

\Cref{fig:eval_pipeline} presents our five-step evaluation pipeline. We executed this pipeline for each real-world dataset used in this evaluation.
The steps are explained below.

\smallskip
\noindent
\textbf{Select events.}
In this step, we select events from the dataset to ensure events have required attributes and remove infrequent case traces. Specifically, every event must have a \emph{case}, \emph{timestamp}, \emph{activity}, and \emph{agent} attribute. Moreover, each selected event must be part of a frequent case variant (a trace within \textit{vff\%} of the most common case traces, where \textit{vff} is the  \textit{variant frequency filter} parameter).

\smallskip
\noindent
\textbf{Identify agent types.}
For the event selection, we identify agent instances using agent attribute values of events and group them into agent types by clustering them based on distances between agent instance DFGs, where an agent instance DFG is constructed from the corresponding agent log (\cref{def:agent_log}). 
For a pair of agents $\pair{a_1}{a_2}$, we calculate the distance between the DFGs of $a_1$ and $a_2$ as: 
$$1 - \max(\nicefrac{\cardinality{\mathit{DF}_{a_1} \cap \mathit{DF}_{a_2}}}{\cardinality{\mathit{DF}_{a_1}}},\nicefrac{\cardinality{\mathit{DF}_{a_1} \cap \mathit{DF}_{a_2}}}{\cardinality{\mathit{DF}_{a_2}}}),$$ 
where $\mathit{DF}_{a_1}$ and $\mathit{DF}_{a_2}$ are the directly-follows relations (sets of edges) in the corresponding DFGs.
Thus, the more one DFG subsumes the directly-follows relations of the other DFG, the more likely the corresponding agent instances belong to the same agent type.
Once the agent types are identified, the agent attribute values of all the events in the input event selection are updated from instances to types. 
For example, this step leads to the identification of three agent types \textsl{a1} (doctor \textsl{d1}), \textsl{a2} (doctors \textsl{d2} and \textsl{d4}), and \textsl{a3} (doctors \textsl{d3} and \textsl{d5}) in the motivating example from \cref{sec:motivating_example}.

\smallskip
\noindent	
\textbf{Discover CM models.}
We refer to a discovery algorithm based on the conventional process mining paradigm as a Conventional Miner (CM) and a model discovered by a CM algorithm as a CM model.
We use two state-of-the-art CM algorithms, Inductive Miner IMf variant (IM)~\cite{leemans2013discovering} and Split Miner (SM)~\cite{Augusto2019}, to construct CM models from the case traces induced by the event selection, that is, traces identified by the case attributes, see \cref{def:event_log}.
Given an event selection $S$, to discover CM models, we use two naming functions: i) activity-only labeling (AOL) defined by $\nu_{\mathit{aol}}=\setbuilder{\pair{e}{\funcCall{e}{\mathit{activity}}}}{e \in S}$ and ii) agent and activity labeling (AAL) defined by $\nu_{\mathit{aal}}=\setbuilder{\pair{e}{\pair{\funcCall{e}{\mathit{agent}}}{\funcCall{e}{\mathit{activity}}}}}{e \in S}$.
Thus, for each event dataset, we construct four CM models.
For example, the model in \cref{fig:soa:running_example_viz} is the IM model constructed using the $\nu_{\mathit{aol}}$ naming function. 
Hence, agent attributes of events were not used to construct the model, and then observable transitions were annotated with agent info.
For each CM algorithm and naming function, we construct ten CM models, one for each configuration of the \emph{threshold} parameter of the algorithm, \emph{noise threshold} for IM and \emph{frequency threshold} for SM (in the range from $0.0$ to $0.9$ in 0.1 increments).
This approach ensures that the discovered CM models show a range of quality characteristics.

\smallskip
\noindent	
\textbf{Discover AM models.}
For the event selection over agent types, we use the Agent Miner (AM) algorithm (refer to \cref{sec:agent_miner}) to discover agent nets, the model of agent interactions, and the MAS net that defines the semantics of the resulting agent systems.
DFG translation to Petri nets (DFG-PN)~\cite{Leemans2019} algorithm is used as an ANDA. 
Inductive Miner (IM)~\cite{leemans2013discovering} and Split Miner (SM)~\cite{Augusto2019} are used as INDAs.
For each configuration of ANDA and INDA, we run AM ten times with ten different parameter pairs $(\mathit{ff}_i,\mathit{th}_i)$, with $i \in \intintervalcc{1}{10}$, $\mathit{ff}_i=i \times 0.1$, and $\mathit{th}_i=1-i \times 0.1$. 
Parameter $\mathit{ff}_i$ is the \emph{activity frequency filter} parameter used by the ANDA. 
Parameter $\mathit{th}_i$ is the \emph{threshold} parameter used by the INDA, \emph{noise threshold} for IM and \emph{frequency threshold} for SM. 
The lower the value of $i$, the more filtered the event selection is and the less the discovered models are fit to the input data.

\smallskip
\noindent
\textbf{Compare models.}
To compare the discovered CM and AM models, we calculate their \emph{recall} and \emph{precision} with respect to the case trace set of the original event selection, and \emph{size} (as the number of nodes and arcs).
To compute precision and recall, we use the entropy-based measures~\cite{PolyvyanyyACGKL20,PolyvyanyySWCM20}.
These measures fulfill all the desired properties for these classes of measures~\cite{DBLP:journals/topnoc/SyringTA19}.
For example, the entropy-based precision measure guarantees that a model that describes fewer traces not in the event log has a better precision score.
For CM models constructed using AOL and AAL naming functions of events, we measure their precision and recall with respect to the event logs that identify events using the corresponding naming functions.
Note that labels of observable transitions of AM models are composed of agent-activity pairs, see, for example, the MAS net in \cref{fig:mas_net}. 
Hence, to compare MAS nets to CM models constructed using the AOL naming function, we also measure precision and recall of MAS nets after rewriting their transition labels to only mention activity names.

\vspace{-3mm}
\subsection{Datasets and Results}
\label{sec:eval_results_synthetic}
\vspace{-2mm}
To evaluate Agent Miner, we used publicly available real-world Business Process Intelligence Challenge (BPIC) datasets and assumed they stem from agent systems.
These datasets are widely used to evaluate conventional process discovery algorithms.
We assumed that the resource attributes of events specify agent instances that triggered them and selected all BPIC datasets that specify events with resource attributes, leading to nine selected datasets.\footnote{\scriptsize%
	BPIC 2011 (\url{https://doi.org/10.4121/uuid:d9769f3d-0ab0-4fb8-803b-0d1120ffcf54}),\\
	BPIC 2012 (\url{https://doi.org/10.4121/uuid:3926db30-f712-4394-aebc-75976070e91f}),\\
	BPIC 2013 (\url{https://doi.org/10.4121/uuid:a7ce5c55-03a7-4583-b855-98b86e1a2b07}),\\
	BPIC 2014 (\url{https://doi.org/10.4121/uuid:3cfa2260-f5c5-44be-afe1-b70d35288d6d}),\\
	BPIC 2015 (\url{https://doi.org/10.4121/uuid:31a308ef-c844-48da-948c-305d167a0ec1}),\\
	BPIC 2017 (\url{https://doi.org/10.4121/uuid:5f3067df-f10b-45da-b98b-86ae4c7a310b}),\\
	BPIC 2018 (\url{https://doi.org/10.4121/uuid:3301445f-95e8-4ff0-98a4-901f1f204972}),\\
	BPIC 2019 (\url{https://doi.org/10.4121/uuid:d06aff4b-79f0-45e6-8ec8-e19730c248f1}), and\\
	BPIC 2020 (\url{https://doi.org/10.4121/uuid:52fb97d4-4588-43c9-9d04-3604d4613b51}).%
}

Initially, we used the \textit{vff}\% parameter of 80\% and completed the evaluation pipeline (cf.~\cref{fig:eval_pipeline}) for datasets BPIC 2012, 2013, 2015 (Municipality~1), 2017, and 2020 (Travel Permit Data).
Due to performance reasons, to process the other datasets, we lowered the \textit{vff}\% parameter to 10\%.
\Cref{fig:bpic_eval_results} summarizes the quality measurements for workflow nets discovered using Inductive Miner and MAS nets constructed by Agent Miner, using activity-only labeling (AOL) of transitions in the discovered nets.
For each dataset, the table lists size, recall, and precision values for the models that scored the lowest size and the greatest precision.

\begin{table}[t]
	\vspace{-3mm}
	\centering
	\caption{\small{Size, recall, and precision of IM nets and MAS nets discovered by Agent Miner (Inductive Miner as INDA) that rely on activity-only labeling of observable transitions.}}
	\label{fig:bpic_eval_results}
	\resizebox{0.85\textwidth}{!}{%
		\begin{tabular}{|c|c|cccccc|cccccc|}
			\hline
			\multicolumn{1}{|l|}{\multirow{3}{*}{\textbf{\begin{tabular}[c]{@{}l@{}}BPIC\\ dataset\end{tabular}}}} & \multirow{3}{*}{\textbf{\begin{tabular}[c]{@{}c@{}}Variant\\ frequency \\ filter (\textit{vff}\%)\end{tabular}}} & \multicolumn{6}{c|}{\textbf{Inductive Miner}} & \multicolumn{6}{c|}{\textbf{Agent Miner (MAS nets)}} \\ \cline{3-14} 
			\multicolumn{1}{|l|}{} &  & \multicolumn{3}{c|}{\textbf{lowest size}} & \multicolumn{3}{c|}{\textbf{greatest precision}} & \multicolumn{3}{c|}{\textbf{lowest size}} & \multicolumn{3}{c|}{\textbf{greatest precision}} \\ \cline{3-14} 
			\multicolumn{1}{|l|}{} &  & \multicolumn{1}{c|}{\textbf{size}} & \multicolumn{1}{c|}{\textbf{recall}} & \multicolumn{1}{c|}{\textbf{prec.}} & \multicolumn{1}{c|}{\textbf{size}} & \multicolumn{1}{c|}{\textbf{recall}} & \textbf{prec.} & \multicolumn{1}{c|}{\textbf{size}} & \multicolumn{1}{c|}{\textbf{recall}} & \multicolumn{1}{c|}{\textbf{prec.}} & \multicolumn{1}{c|}{\textbf{size}} & \multicolumn{1}{c|}{\textbf{recall}} & \textbf{prec.} \\ \hline
			2011 & 10\% & \multicolumn{1}{c|}{592} & \multicolumn{1}{c|}{0.72} & \multicolumn{1}{c|}{\textbf{0.41}} & \multicolumn{1}{c|}{592} & \multicolumn{1}{c|}{0.72} & \textbf{0.41} & \multicolumn{1}{c|}{\textbf{466}} & \multicolumn{1}{c|}{\textbf{0.92}} & \multicolumn{1}{c|}{\textbf{0.35}} & \multicolumn{1}{c|}{\textbf{466}} & \multicolumn{1}{c|}{\textbf{0.92}} & 0.35 \\ \hline
			2012 & 80\% & \multicolumn{1}{c|}{420} & \multicolumn{1}{c|}{\textbf{1.00}} & \multicolumn{1}{c|}{0.06} & \multicolumn{1}{c|}{454} & \multicolumn{1}{c|}{\textbf{0.88}} & \textbf{0.32} & \multicolumn{1}{c|}{\textbf{333}} & \multicolumn{1}{c|}{0.80} & \multicolumn{1}{c|}{\textbf{0.18}} & \multicolumn{1}{c|}{\textbf{333}} & \multicolumn{1}{c|}{0.80} & 0.18 \\ \hline
			2013 & 80\% & \multicolumn{1}{c|}{\textbf{54}} & \multicolumn{1}{c|}{\textbf{0.70}} & \multicolumn{1}{c|}{0.53} & \multicolumn{1}{c|}{105} & \multicolumn{1}{c|}{\textbf{0.78}} & 0.62 & \multicolumn{1}{c|}{69} & \multicolumn{1}{c|}{0.62} & \multicolumn{1}{c|}{0.64} & \multicolumn{1}{c|}{\textbf{69}} & \multicolumn{1}{c|}{0.62} & \textbf{0.64} \\ \hline
			2014 & 10\% & \multicolumn{1}{c|}{\textbf{229}} & \multicolumn{1}{c|}{\textbf{0.85}} & \multicolumn{1}{c|}{0.28} & \multicolumn{1}{c|}{335} & \multicolumn{1}{c|}{0.55} & \textbf{0.45} & \multicolumn{1}{c|}{231} & \multicolumn{1}{c|}{0.78} & \multicolumn{1}{c|}{\textbf{0.36}} & \multicolumn{1}{c|}{\textbf{231}} & \multicolumn{1}{c|}{\textbf{0.78}} & 0.36 \\ \hline
			2015 & 80\% & \multicolumn{1}{c|}{199} & \multicolumn{1}{c|}{\textbf{0.95}} & \multicolumn{1}{c|}{\textbf{0.35}} & \multicolumn{1}{c|}{199} & \multicolumn{1}{c|}{\textbf{0.95}} & \textbf{0.35} & \multicolumn{1}{c|}{\textbf{122}} & \multicolumn{1}{c|}{\textbf{0.95}} & \multicolumn{1}{c|}{\textbf{0.34}} & \multicolumn{1}{c|}{\textbf{122}} & \multicolumn{1}{c|}{\textbf{0.95}} & 0.34 \\ \hline
			2017 & 80\% & \multicolumn{1}{c|}{\textbf{216}} & \multicolumn{1}{c|}{\textbf{0.85}} & \multicolumn{1}{c|}{\textbf{0.22}} & \multicolumn{1}{c|}{\textbf{241}} & \multicolumn{1}{c|}{0.91} & \textbf{0.26} & \multicolumn{1}{c|}{340} & \multicolumn{1}{c|}{0.76} & \multicolumn{1}{c|}{0.15} & \multicolumn{1}{c|}{597} & \multicolumn{1}{c|}{\textbf{0.94}} & 0.15 \\ \hline
			2018 & 10\% & \multicolumn{1}{c|}{\textbf{350}} & \multicolumn{1}{c|}{\textbf{0.88}} & \multicolumn{1}{c|}{\textbf{0.13}} & \multicolumn{1}{c|}{\textbf{350}} & \multicolumn{1}{c|}{\textbf{0.88}} & \textbf{0.13} & \multicolumn{1}{c|}{535} & \multicolumn{1}{c|}{0.85} & \multicolumn{1}{c|}{0.09} & \multicolumn{1}{c|}{535} & \multicolumn{1}{c|}{0.86} & 0.09 \\ \hline
			2019 & 10\% & \multicolumn{1}{c|}{270} & \multicolumn{1}{c|}{0.61} & \multicolumn{1}{c|}{0.24} & \multicolumn{1}{c|}{375} & \multicolumn{1}{c|}{0.54} & \textbf{0.46} & \multicolumn{1}{c|}{\textbf{191}} & \multicolumn{1}{c|}{\textbf{0.65}} & \multicolumn{1}{c|}{\textbf{0.34}} & \multicolumn{1}{c|}{\textbf{191}} & \multicolumn{1}{c|}{\textbf{0.65}} & 0.34 \\ \hline
			2020 & 80\% & \multicolumn{1}{c|}{204} & \multicolumn{1}{c|}{0.72} & \multicolumn{1}{c|}{0.18} & \multicolumn{1}{c|}{220} & \multicolumn{1}{c|}{0.69} & \textbf{0.25} & \multicolumn{1}{c|}{\textbf{200}} & \multicolumn{1}{c|}{\textbf{0.75}} & \multicolumn{1}{c|}{\textbf{0.19}} & \multicolumn{1}{c|}{\textbf{200}} & \multicolumn{1}{c|}{\textbf{0.75}} & 0.19 \\ \hline
		\end{tabular}%
	}
	\vspace{-3mm}
\end{table}

\noindent
These results confirm that the quality of the MAS nets discovered by Agent Miner is comparable to the quality of CM models.
This observation is remarkable for at least two reasons.
First, as stated above, we had no background knowledge of whether the datasets stem from agent-driven business processes.
Still, for most datasets, we discovered interesting (in terms of size, recall, and precision) MAS nets.
Second, in addition to high-quality MAS nets, Agent Miner constructs agent nets and i-nets that can be used as a starting point for analysis and improvement of ways the process participants work individually and together.

\begin{figure}[t]
	\centering
	\begin{subfigure}[b]{0.325\textwidth}
		\centering
		\includegraphics[height=2.0cm,trim=5mm 5mm 5mm 5mm]{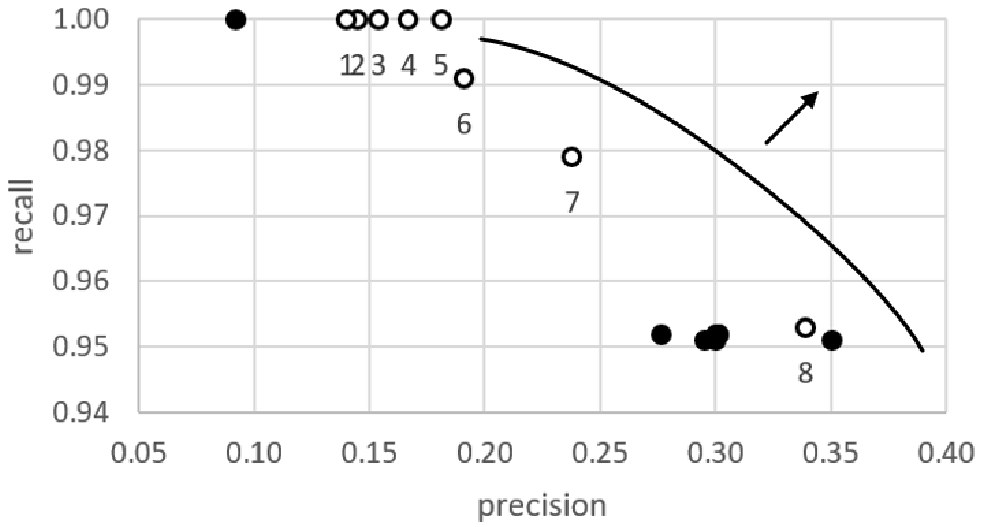}
		\caption{\small{recall/prec. trade-off}}
		\label{bpic2015_pareto_fronts_rp}
	\end{subfigure}
	\begin{subfigure}[b]{0.325\textwidth}
		\centering
		\includegraphics[height=2.0cm,trim=5mm 5mm 5mm 5mm]{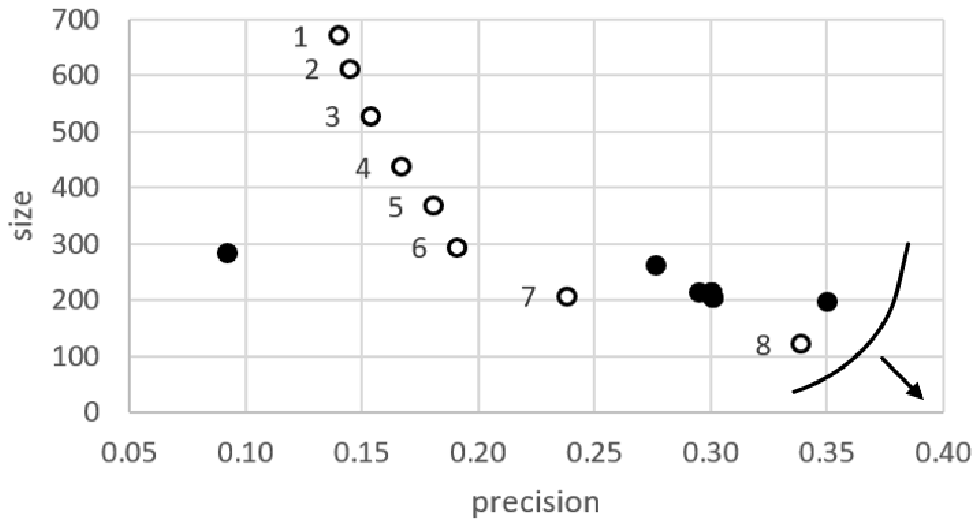}
		\caption{\small{size/prec. trade-off}}
		\label{bpic2015_pareto_fronts_sp}
	\end{subfigure}
	\begin{subfigure}[b]{0.325\textwidth}
		\centering
		\includegraphics[height=2.0cm,trim=5mm 5mm 5mm 5mm]{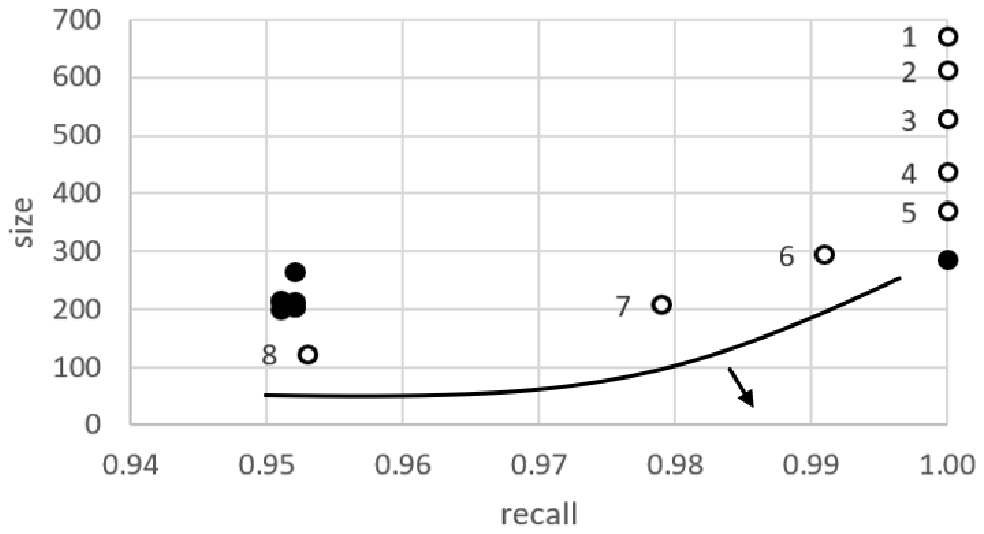}
		\caption{\small{size/recall trade-off}}
		\label{bpic2015_pareto_fronts_sr}
	\end{subfigure}
	\vspace{-2mm}
	\caption{\small{BPIC 2015 Pareto fronts: IM--AOL nets (``$\bullet$'') and AM--AOL MAS nets (``$\circ$'').}}
	\label{fig:bpic2015_pareto_fronts}
\end{figure}

\begin{figure}[t]
	\centering
	\begin{subfigure}[b]{0.32\textwidth}
		\centering
		\includegraphics[height=1.67cm,trim=5mm 5mm 5mm 5mm]{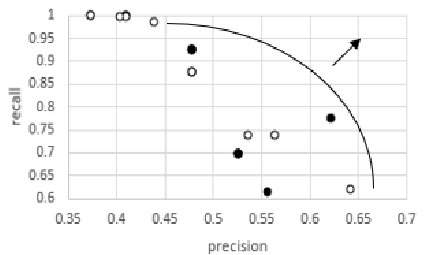}
		\vspace{3mm}
		\caption{\small{recall/prec. trade-off}}
		\label{bpic2013_pareto_fronts_rp}
	\end{subfigure}
	\begin{subfigure}[b]{0.32\textwidth}
		\centering
		\includegraphics[height=1.73cm,trim=5mm 5mm 5mm 5mm]{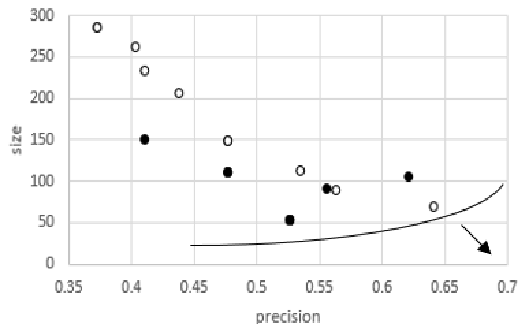}
		\vspace{3mm}
		\caption{\small{size/prec. trade-off}}
		\label{bpic2013_pareto_fronts_sp}
	\end{subfigure}
	\begin{subfigure}[b]{0.32\textwidth}
		\centering
		\includegraphics[height=1.64cm,trim=5mm 5mm 5mm 5mm]{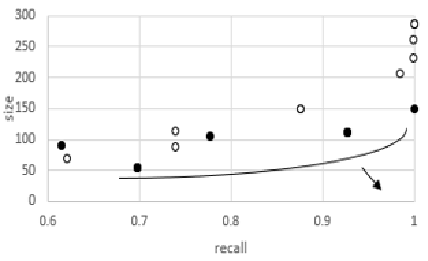}
		\vspace{3mm}
		\caption{\small{size/recall trade-off}}
		\label{bpic2013_pareto_fronts_sr}
	\end{subfigure}
	\vspace{-2mm}
	\caption{\small{BPIC 2013 Pareto fronts: IM--AOL nets (``$\bullet$'') and AM--AOL MAS nets (``$\circ$'').}}
	\label{fig:bpic2013_pareto_fronts}
	\vspace{-7mm}
\end{figure}

To support the above conclusions, \cref{fig:bpic2015_pareto_fronts} shows three types of trade-offs as two-dimensional Pareto fronts for the nets discovered from the BPIC 2015 dataset.
Each point in the plots denotes two quality measurements for one net generated by the evaluation pipeline.
The Pareto fronts indicate the nets with better quality measurements. 
The MAS nets discovered by Agent Miner complement workflow nets discovered by Inductive Miner to result in more saturated Pareto fronts, that is, fronts with more models of interesting qualities.
For the BPIC15 dataset, MAS nets 1 to 8 discovered by Agent Miner belong to the recall/precision Pareto front and demonstrate combinations of these quality measurements better than most CM models discovered by Inductive Miner. 
MAS net 8 belongs to the size/precision Pareto front and is better in terms of size/precision than most of the CM models. 
MAS nets 1 to 5, 7, and 8 demonstrate a better combination of size/recall measurements than almost all CM models.
Overall, the Pareto fronts contain more AM models than CM models, except in the size/precision case, when the front is represented by one AM model and one CM model.
Similar to the BPIC 2015 results, the Pareto fronts for the BPIC 2013 dataset, presented in \cref{fig:bpic2013_pareto_fronts}, include points for the MAS nets.
The Pareto front plots for all the datasets are included in the evaluation results~\cite{Tour2023data}.

Agent Miner uses an additional event attribute that specifies the \emph{agent} that triggered the event.
Hence, it is reasonable to expect it to construct comparable or better models than conventional algorithms, which do not require this attribute.
To obtain generalizable conclusions, for all the datasets, we analyzed Pareto fronts and performed paired samples t-tests to establish whether CM and AM models are of the same qualities, for CM models discovered using both activity-only labeling (AOL) and agent and activity labeling (AAL).
When measuring precision and recall of AOL and AAL models, both AM and CM models, events in the logs were identified correspondingly. 
The null hypotheses used in the t-tests are such that means of quality measurements for AM and CM models are equal.
\cref{tbl:bpic_eval_pareto} summarizes comparisons of Pareto fronts for IM nets and AM nets (IM as INDA), while \cref{tbl:bpic_eval_test} shows the results of the tests, where results are for size (s), recall (r) and precision (p) measurements.
In \cref{tbl:bpic_eval_pareto}, AM, CM, and AM\&CM entries stand for situations when the Pareto front is composed of AM only, CM only, or AM and CM models, respectively.
In \cref{tbl:bpic_eval_test}, for the significance level of 0.05, AM, CM, and AM\&CM entries stand for situations when AM models are significantly better, CM models are significantly better, and the null hypothesis was not rejected, respectively.
We accept the AM and AM\&CM entries as situations when Agent Miner provides additional value to CM models and highlight them in bold (the majority of the entries).

\begin{table}[t]
	\vspace{-4mm}
	\centering
	\makebox[0pt][c]{\parbox{1.0\textwidth}{%
			\begin{minipage}{0.49\textwidth}
				\caption{\small{Comparison of Pareto fronts.}}
				\label{tbl:bpic_eval_pareto}
				\resizebox{1.0\textwidth}{!}{%
					\begin{tabular}{|
							>{\columncolor[HTML]{EFEFEF}}c |ccc|ccc|}
						\hline
						\multicolumn{1}{|l|}{\cellcolor[HTML]{EFEFEF}}                                                                                  & \multicolumn{3}{c|}{\cellcolor[HTML]{EFEFEF}\textbf{AM vs CM fronts (AOL)}}                                                                                  & \multicolumn{3}{c|}{\cellcolor[HTML]{EFEFEF}\textbf{AM vs CM fronts (ALL)}}                                                                                  \\ \cline{2-7} 
						\multicolumn{1}{|l|}{\multirow{-2}{*}{\cellcolor[HTML]{EFEFEF}\textbf{\begin{tabular}[c]{@{}l@{}}BPIC\\ data\end{tabular}}}} & \multicolumn{1}{c|}{\cellcolor[HTML]{EFEFEF}\textbf{r/p}} & \multicolumn{1}{c|}{\cellcolor[HTML]{EFEFEF}\textbf{s/p}} & \cellcolor[HTML]{EFEFEF}\textbf{s/r} & \multicolumn{1}{c|}{\cellcolor[HTML]{EFEFEF}\textbf{r/p}} & \multicolumn{1}{c|}{\cellcolor[HTML]{EFEFEF}\textbf{s/p}} & \cellcolor[HTML]{EFEFEF}\textbf{s/r} \\ 
						\hline
						\hline
						\textbf{2011}                                                                                                                   & \multicolumn{1}{c|}{\textbf{AM}}                          & \multicolumn{1}{c|}{\textbf{AM\&CM}}                      & \textbf{AM}                          & \multicolumn{1}{c|}{\textbf{AM\&CM}}                      & \multicolumn{1}{c|}{CM}                                   & \textbf{AM\&CM}                      \\ \hline
						\textbf{2012}                                                                                                                   & \multicolumn{1}{c|}{\textbf{AM\&CM}}                      & \multicolumn{1}{c|}{\textbf{AM\&CM}}                      & \textbf{AM\&CM}                      & \multicolumn{1}{c|}{\textbf{AM\&CM}}                      & \multicolumn{1}{c|}{\textbf{AM\&CM}}                      & \textbf{AM}                          \\ \hline
						\textbf{2013}                                                                                                                   & \multicolumn{1}{c|}{\textbf{AM\&CM}}                      & \multicolumn{1}{c|}{\textbf{AM\&CM}}                      & \textbf{AM\&CM}                      & \multicolumn{1}{c|}{\textbf{AM\&CM}}                      & \multicolumn{1}{c|}{\textbf{AM\&CM}}                      & \textbf{AM\&CM}                      \\ \hline
						\textbf{2014}                                                                                                                   & \multicolumn{1}{c|}{\textbf{AM\&CM}}                      & \multicolumn{1}{c|}{\textbf{AM\&CM}}                      & CM                                   & \multicolumn{1}{c|}{\textbf{AM\&CM}}                      & \multicolumn{1}{c|}{\textbf{AM\&CM}}                      & \textbf{AM\&CM}                      \\ \hline
						\textbf{2015}                                                                                                                   & \multicolumn{1}{c|}{\textbf{AM\&CM}}                      & \multicolumn{1}{c|}{\textbf{AM\&CM}}                      & \textbf{AM\&CM}                      & \multicolumn{1}{c|}{\textbf{AM\&CM}}                      & \multicolumn{1}{c|}{CM}                                   & CM                                   \\ \hline
						\textbf{2017}                                                                                                                   & \multicolumn{1}{c|}{\textbf{AM\&CM}}                      & \multicolumn{1}{c|}{CM}                                   & CM                                   & \multicolumn{1}{c|}{\textbf{AM\&CM}}                      & \multicolumn{1}{c|}{\textbf{AM\&CM}}                      & CM                                   \\ \hline
						\textbf{2018}                                                                                                                   & \multicolumn{1}{c|}{CM}                                   & \multicolumn{1}{c|}{CM}                                   & CM                                   & \multicolumn{1}{c|}{CM}                                   & \multicolumn{1}{c|}{CM}                                   & CM                                   \\ \hline
						\textbf{2019}                                                                                                                   & \multicolumn{1}{c|}{\textbf{AM\&CM}}                      & \multicolumn{1}{c|}{\textbf{AM\&CM}}                      & \textbf{AM\&CM}                      & \multicolumn{1}{c|}{CM}                                   & \multicolumn{1}{c|}{CM}                                   & \textbf{AM\&CM}                      \\ \hline
						\textbf{2020}                                                                                                                   & \multicolumn{1}{c|}{\textbf{AM\&CM}}                      & \multicolumn{1}{c|}{\textbf{AM\&CM}}                      & \textbf{AM\&CM}                      & \multicolumn{1}{c|}{\textbf{AM\&CM}}                      & \multicolumn{1}{c|}{\textbf{AM\&CM}}                      & \textbf{AM\&CM}                      \\ \hline
					\end{tabular}
				}
			\end{minipage}%
			\hfill 
			\begin{minipage}{0.495\textwidth}
				\caption{\small{Results of t-tests.}}
				\label{tbl:bpic_eval_test}
				\resizebox{1.0\textwidth}{!}{%
					\begin{tabular}{|
							>{\columncolor[HTML]{EFEFEF}}c |ccc|ccc|}
						\hline
						\multicolumn{1}{|l|}{\cellcolor[HTML]{EFEFEF}}                                                                                  & \multicolumn{3}{c|}{\cellcolor[HTML]{EFEFEF}\textbf{AM vs CM tests (AOL)}}                                                                             & \multicolumn{3}{c|}{\cellcolor[HTML]{EFEFEF}\textbf{AM vs CM tests (ALL)}}                                                                             \\ \cline{2-7} 
						\multicolumn{1}{|l|}{\multirow{-2}{*}{\cellcolor[HTML]{EFEFEF}\textbf{\begin{tabular}[c]{@{}l@{}}BPIC\\ data\end{tabular}}}} & \multicolumn{1}{c|}{\cellcolor[HTML]{EFEFEF}\,\,\,\,\,\,\,\,\,\,\,\,\,\,\textbf{s}\,\,\,\,\,\,\,\,\,\,\,\,\,\,} & \multicolumn{1}{c|}{\cellcolor[HTML]{EFEFEF}\textbf{r}} & \cellcolor[HTML]{EFEFEF}\textbf{p} & \multicolumn{1}{c|}{\cellcolor[HTML]{EFEFEF}\textbf{s}} & \multicolumn{1}{c|}{\cellcolor[HTML]{EFEFEF}\textbf{r}} & \cellcolor[HTML]{EFEFEF}\textbf{p} \\ 
						\hline
						\hline
						\textbf{2011}                                                                                                                   & \multicolumn{1}{c|}{CM}                                 & \multicolumn{1}{c|}{CM}                                 & CM                                 & \multicolumn{1}{c|}{CM}                                 & \multicolumn{1}{c|}{\textbf{AM}}                        & \textbf{AM\&CM}                    \\ \hline
						\textbf{2012}                                                                                                                   & \multicolumn{1}{c|}{CM}                                 & \multicolumn{1}{c|}{CM}                                 & \textbf{AM\&CM}                    & \multicolumn{1}{c|}{\textbf{AM\&CM}}                    & \multicolumn{1}{c|}{\textbf{AM}}                        & \textbf{AM\&CM}                    \\ \hline
						\textbf{2013}                                                                                                                   & \multicolumn{1}{c|}{CM}                                 & \multicolumn{1}{c|}{\textbf{AM\&CM}}                    & \textbf{AM\&CM}                    & \multicolumn{1}{c|}{\textbf{AM\&CM}}                    & \multicolumn{1}{c|}{\textbf{AM}}                        & \textbf{AM\&CM}                    \\ \hline
						\textbf{2014}                                                                                                                   & \multicolumn{1}{c|}{CM}                                 & \multicolumn{1}{c|}{CM}                                 & \textbf{AM\&CM}                    & \multicolumn{1}{c|}{CM}                                 & \multicolumn{1}{c|}{\textbf{AM}}                        & \textbf{AM\&CM}                    \\ \hline
						\textbf{2015}                                                                                                                   & \multicolumn{1}{c|}{CM}                                 & \multicolumn{1}{c|}{\textbf{AM\&CM}}                    & \textbf{AM\&CM}                    & \multicolumn{1}{c|}{CM}                                 & \multicolumn{1}{c|}{\textbf{AM}}                        & \textbf{AM\&CM}                    \\ \hline
						\textbf{2017}                                                                                                                   & \multicolumn{1}{c|}{CM}                                 & \multicolumn{1}{c|}{\textbf{AM\&CM}}                    & \textbf{AM\&CM}                    & \multicolumn{1}{c|}{\textbf{AM\&CM}}                    & \multicolumn{1}{c|}{\textbf{AM\&CM}}                    & \textbf{AM\&CM}                    \\ \hline
						\textbf{2018}                                                                                                                   & \multicolumn{1}{c|}{CM}                                 & \multicolumn{1}{c|}{CM}                                 & \textbf{AM\&CM}                    & \multicolumn{1}{c|}{\textbf{AM\&CM}}                    & \multicolumn{1}{c|}{\textbf{AM\&CM}}                    & \textbf{AM\&CM}                    \\ \hline
						\textbf{2019}                                                                                                                   & \multicolumn{1}{c|}{CM}                                 & \multicolumn{1}{c|}{CM}                                 & \textbf{AM}                        & \multicolumn{1}{c|}{CM}                                 & \multicolumn{1}{c|}{\textbf{AM}}                        & \textbf{AM\&CM}                    \\ \hline
						\textbf{2020}                                                                                                                   & \multicolumn{1}{c|}{CM}                                 & \multicolumn{1}{c|}{CM}                                 & \textbf{AM\&CM}                    & \multicolumn{1}{c|}{CM}                                 & \multicolumn{1}{c|}{\textbf{AM}}                        & CM                                 \\ \hline
					\end{tabular}
				}
			\end{minipage}%
	}}
	\vspace{-4mm}
\end{table}

The results confirm that Agent Miner discovers interesting models to complement models constructed by conventional discovery algorithms that also rely on event attributes that specify agents that triggered the events and describe this information in the constructed models.
The results for Split Miner, CM models, and AM models (Split Miner as INDA), demonstrate similar conclusions as for IM models presented above.

\vspace{-3mm}
\section{Related Work}
\label{sec:related:work}
\vspace{-2mm}

This section reviews
business process modeling with agents, process discovery
addressing agents,
and agent system modeling in a broad context. 

\textbf{Agent Modeling in Business Process Management.} 
The relationship between business process management
and agent-based modeling was first explored in the `90s by 
Jennings et al.~\cite{jennings1996agent}. 
In an extension of their work, dubbed ADEPT, the authors 
propose to 
model a business process as a 
negotiation system between agents, 
similar to our interaction nets~\cite{jennings1998adept}. However,
their model of negotiating agents is conceptual in its nature, while 
our model is formal and executable. 
Several authors advocate for an 
agent-based perspective on business process modeling~\cite{halavska2018there,sulis2017agent}. In these works, the process
is seen as such composed of interacting agents~\cite{jennings1996agent}. However, they do not 
provide automated discovery of models
from data. 

\textbf{Process Discovery.} 
Traditional process discovery 
techniques assume a case perspective when learning models 
from data while often ignoring additional perspectives such as resources.
Several process discovery techniques extended traditional methods beyond the case perspective. 
Rozinat et al.~\cite{rozinat2009discovering} propose a multi-perspective approach 
for mining simulation models from event data that
includes the resource perspective. 
The resulting models are executable and can be used for performance analysis of the underlying system. However, resources are considered static entities and not active agents. Van der Aalst et al.~\cite{van2010business} address the modeling of behavior and availability of resources. Yet, resources play a secondary role, with cases
being the dominant perspective that defines business processes. Klijn et al.~\cite{fahland2022pqmi}
develop a technique for querying event logs to uncover 
interactions between process entities. Yet, they do not provide a formal model
that can be evaluated for its correctness or goodness-of-fit. 
Discovering functional architecture models (FAM)
that comprise interacting
modules that internally perform various activities was proposed in~\cite{van2015discovery}. Unlike the modules in FAMs, our agents
are dynamic, decentralized, and may interact not only with other agents, but also with the environment. 
Fettke and Reisig presented an approach to system mining called Heraklit~\cite{FettkeR22}.
Heraklit proceeds by constructing distributed runs of participating agents and then combining them into the overall system, whereas Agent Miner constructs agent nets and an additional net that explicitly describes agents' interactions.

Tour et al.~\cite{tour2021agent} have shown that by shifting process mining paradigm from case-based to agent-based, one can discover less complex models of business processes. Within this agent paradigm, Nesterov et al.~\cite{nesterov2022discovering} have recently proposed a process discovery solution. Their algorithm constructs sound generalized workflow nets that capture agents' behavior. In contrast to our approach, Nesterov et al. assume to know the interaction patterns between agents and aim to implement this given interaction pattern, while we discover the interactions from the data. Moreover, their algorithm can only discover 
pre-defined workflow patterns, while Agent Miner is generic and can model any pattern its ANDA component can discover. Furthermore, we evaluated Agent Miner over real-life logs, while Nesterov et al. tested their approach over synthetically generated data. 

Our approach can also be viewed as a log-decomposition process discovery approach~\cite{van2015process}.
Such techniques propose to localize the event log to discover different models tailored to the data in these local logs. In these techniques, the log decomposition is usually driven by the case attribute, whereas we propose an agent-driven log decomposition.

\textbf{Multi-Agent System Modeling.} Multi-agent systems (MASs) have been studied extensively in the past; cf.~\cite{dorri2018multi} for a recent overview. Here, we focus on approaches that model multi-agent
systems using Petri nets, our formalism of choice. 
A seminal paper by Moldt et al.~\cite{moldt1997multi} proposed to use colored Petri nets (CPNs) to model agent systems. 
Their model captures three crucial components in a MAS: communication, independence (between agents), and intelligence. 
However, given the richness of CPNs, formal results on 
correctness of constructed models were not obtain. Celaya et al.~\cite{celaya2007modeling} model a multi-agent system using elementary
Petri nets to capture interactions
between agents and use Petri net analysis to ensure the models are deadlock-free. In our work, we discover MAS models 
from event data rather than relying on human expertise.

\vspace{-2mm}
\section{Conclusion}
\label{sec:conclusion}
\vspace{-2mm}

This paper presents and evaluates Agent Miner, a divide-and-conquer algorithm for discovering models of agents and their interactions from event data. The algorithm ``divides’’ the input data into special parts and then ``conquers’’ the parts using conventional process discovery algorithms. The constructed agent and interaction models provide a new, modular perspective on the data, suitable for analyzing process participants and their interactions. These artifacts can be integrated into a model that describes process control flow. Such integrated models are often smaller and represent the event data more faithfully than corresponding process models constructed using conventional discovery algorithms. The configuration of Agent Miner used in the evaluation reported in this paper ensures that the obtained integrated models are safe and sound.

Agent Miner has several limitations representing areas of interest for future work.
First, the interaction logs are constructed by taking the first event in each agent trace of each agent.
This approach ignores the information on the duration of agent activities and interactions.
For instance, information on the durations of activities between interactions can be obtained by considering the first and the last event from agent traces between those interactions.
Consequently, one can apply lifecycle-aware process discovery~\cite{Leemans2015} or queue mining techniques~\cite{Senderovich2019}  to infer agent interactions.
Next, Agent Miner associates each event with only one agent. 
One can relax this limitation and study the effects of multiple agents sharing the same event.
Finally, the evaluation approach used in this article is limited to the traditional model quality measures used in process mining.
The use of new agent-specific quality measures for discovered models may highlight additional benefits of Agent Miner and other agent system mining algorithms in the context of agent-based business process management.

\smallskip%
\noindent%
{\textbf{Acknowledgments.} 
	Andrei Tour was supported via an ``Australian Government Research Training Program Scholarship.''
	Artem Polyvyanyy was in part supported by the Australian Research Council project DP220101516.}

	
	
	\vspace{-3mm}
	\bibliography{main}
	
\end{document}